\documentclass[12pt]{iopart}
\usepackage[utf8]{inputenc}
\usepackage{graphicx}

\usepackage{times}
\usepackage{natbib}

\usepackage{bm}

\makeatletter
\def\lesssim{\mathrel{\mathpalette\vereq<}}
\def\vereq#1#2{\lower0.8ex\vbox{\baselineskip1.5pt \lineskip1.5pt
\ialign{$\m@th#1\hfill##\hfil$\crcr#2\crcr\sim\crcr}}}
\def\gtrsim{\mathrel{\mathpalette\vereq>}}
\makeatother

\newcommand{\BdG}{BdG}
\newcommand{\Lop}{L}

\begin{document} 
\sloppy

\title{Excitations at the border of a condensate}
\author{
Abdoulaye Diallo$^{1,2}$ and Carsten Henkel$^{1}$
\footnote[5]{henkel@uni-potsdam.de}
}
\address{$^{1}$\ Institute of Physics and Astronomy, University of Potsdam, 
Karl-Liebknecht-Str. 24/25, 14476 Potsdam, Germany}
\address{$^{2}$\ 2700 Rolido Drive, Houston TX 77063, USA}

\begin{abstract}
We solve the Bogoliubov--de Gennes equations for an inhomogeneous
condensate in the vicinity of a linear turning point. A stable integration
scheme is developed using a transformation into an adiabatic basis.
We identify boundary modes trapped in a potential whose shape is
similar to a Hartree-Fock mean-field treatment. These modes are 
non-resonantly excited when bulk modes reflect at the turning point
and contribute significantly to the spectrum of local density fluctuations. 
\end{abstract}


\submitto{\jpb}

\section*{Introduction}

The achievement of Bose-Einstein condensation in ultra-cold trapped 
atomic gases \citep{PitaevskiiStringari} has provided experimentalists
with a `direct look' at quantum mechanical wave functions. In addition,
the atom-atom interactions that become relevant despite the low densities,
lead to a nonlinear wave mechanics of degenerate Bose gases, as described
by the celebrated Gross-Pitaevskii equation at the mean field level
\citep{Gross61,Pitaevskii61}, see Eq.(\ref{eq:GPE-3D}) below.
Nonlinearity brings in qualitatively
new features in inhomogeneous systems, for example: 
by neglecting the kinetic energy (second derivative), one gets
a nontrivial solution with a fixed amplitude, the so-called
Thomas-Fermi condensate. This approximation breaks down in the vicinity 
of a turning point, and the condensate's kinetic energy acquires 
logarithmic corrections \citep{Dalfovo96,Fetter98a}. One has to deal 
with a nonlinear boundary layer problem, similar to the Ginzburg-Landau
description of the surface of a superconductor~\citep{Landau9}
that leads to the distinction between type~I and~II superconductors.

We address in this paper the wave mechanics of elementary excitations 
around the Gross-Pitaevskii equation 
by focusing on a typical turning point where the trapping potential is
approximately linear. This situation is of course well known for the linear
Schr\"odinger equation: it leads to an Airy function and the famous
$\pi/4$ phase when semiclassical 
wave functions 
{}(Wenzel-Kramers-Brillouin, WKB) 
are matched on both sides of the turning 
point \citep{Langer37,MessiahI}. In the nonlinear case, one is dealing with
two coupled wave functions or Bogoliubov--de Gennes (\BdG) modes
$u$ and $v$.
This complicates the semiclassical analysis and has led to
modified WKB techniques~\citep{Hyougouchi02}. A
straightforward numerical approach is impossible because the
higher (fourth) order of the wave equation actually generates an instability.
One of the motivations of the present analysis is to provide a
robust scheme for the \BdG{} modes that can be
used as a stepping stone for inhomogeneous low-dimensional Bose gases
at finite temperature. Indeed, in this case, thermally excited
modes give a dominant contribution in the infrared and enforce
the introduction of the quasi-condensate 
concept~\citep{Kagan00,Andersen02c,Mora03}. 
The Bogoliubov modes that we derive here capture the role of spatial
coherence (delocalised waves) and may provide 
a quantitative assessment of the physics beyond the 
local density approximation. 
Indeed, we find that spatial gradients of the condensate density
play a key role for the elementary excitations in the border region.

The paper is organised as follows. We recall the mean-field theory
for the elementary excitations of an inhomogeneous degenerate Bose gas 
and formulate the boundary conditions on both sides of
the position where the chemical potential crosses a linear(ised) 
trapping potential (Sec.\ref{s:model}). In Sec.\ref{s:uv-modes},
the \BdG{} equations are solved approximately with the help of
an adiabatic basis that generalises the transformation to
density and phase modes $u \pm v$. We discuss in particular
the appearance of `trapped modes' near the condensate boundary.
The consequences for physical observables like the condensate
depletion, the average thermal density
and its fluctuations are illustrated in Sec.\ref{s:applications}.
In a companion paper~\citep{Diallo-phase}, we analyse the 
correction to the WKB (Langer) phase at the nonlinear turning
point and its role for the spectral density of elementary modes.

\section{Model}
\label{s:model}

Interacting Bose gases at low temperatures are quite successfully
described by a mean-field theory provided most of the particles occupy 
the condensate mode. This mode then solves a non-linear Schr\"odinger
equation, also known as the Gross--Pitaevskii equation (GPE):
\begin{equation}
- \frac{ \hbar^2 }{ 2 m } \nabla^2 \psi +
V \psi + \frac{ 4\pi\hbar^2 a_s }{ m } |\psi|^2 \psi =  \mu \psi
\label{eq:GPE-3D}
\end{equation}
This is the stationary form of the GPE, with the eigenvalue $\mu$
called the chemical potential. The (positive) scattering length $a_s$ 
specifies
the density-dependence of the inter-particle interactions at the
mean-field level, and $V = V( {\bf r})$ is an external potential.
In this paper, we focus on a quasi-one-dimensional trap and replace
the interaction term by an 
effective interaction strength proportional to $a_s$ and the
transverse confinement. In addition, we focus on the spatial region 
where the potential can be linearized, more specifically in the vicinity 
of a turning point: $V( z ) \approx \mu - F z$. By shifting the
$z$-coordinate, the chemical potential drops out of the GPE. 
With the proper
choice of units (see Table~\ref{t:units}), the GPE finally takes a
universal form~\citep{Dalfovo96}, also recognisable as the second
Painlev\'e transcendent~\citep{Ablowitz77}. 
\begin{equation}
- \frac{ {\rm d}^2 \phi }{ {\rm d}z^2 } - z \phi + 
|\phi|^2 \phi = 0
	\label{eq:GPE}
\end{equation}
The linearisation
around the mean field leads to the Bogoliubov--de Gennes equations
that in the same units can be written as
\begin{eqnarray}
- \frac{ {\rm d}^2 u }{ {\rm d}z^2 } - z u + 2 |\phi|^2 u
+ \phi^2 v^* &=& E u
\nonumber\\
- \frac{ {\rm d}^2 v }{ {\rm d}z^2 } - z v + 2 |\phi|^2 v
+ \phi^{2} u^* &=& - E v
	\label{eq:BdG}
\end{eqnarray}
where $E \ge 0$ is the frequency (energy) of the elementary excitation,
measured relative to the chemical potential.
We fix the phases of $\phi, u, v$ to be real, choosing positive $\phi$.

\begin{table}[bth]
\begin{center}
\fbox{
\begin{tabular}{lllll}
System 
	& Length 
	& Temperature 
	& Frequency
	& Density$^*$
\\
\hline
\\[-2.8ex]
	& $\displaystyle \ell = \hbar^{2/3} / ( 2 m F )^{1/3}$
	& $\displaystyle F \ell / k_B $
	& $\displaystyle F \ell / (2\pi\hbar) $
	& $\displaystyle F \ell / g$
\\
1D, gravity 
	& $0.3\,\mu{\rm m}$
	& $31\,{\rm nK}$
	& $640\,{\rm Hz}$
	& $6.4 / \mu{\rm m}$
\\  
1D, $100\,\mu{\rm m}$ length,
$10\,{\rm Hz}$ trap
	& $1.1\,\mu{\rm m}$
	& $2.3\,{\rm nK}$
	& $48\,{\rm Hz}$
	& $0.47 / \mu{\rm m}$
\\
3D, $100\,\mu{\rm m}$ diam,
$30\,{\rm Hz}$ trap
	& $0.53\,\mu{\rm m}$
	& $9.9\,{\rm nK}$
	& $210\,{\rm Hz}$
	& $28/\mu{\rm m}^3$
\end{tabular}}
\end{center}
\caption[]{Natural units for the Bogoliubov--de Gennes (\BdG)
equations in a linear potential $V = - F z$. The interaction constant
in the quasi-1D geometries (first and second line) is $g 
= 2 \hbar \omega_\perp a_{\rm s}$ with transverse trapping 
frequency $\omega_\perp/2\pi = 10\,{\rm kHz}$ and
s-wave scattering length $a_{\rm s} = 95\,a_0$ 
for Rb87~\citep{Egorov11}. The trapped systems are considered in a
harmonic confinement, the potential being linearised at the
Thomas-Fermi radius.\\
\hspace*{\parindent}
${}^*$The density scale for the 3D trap
is taken as $1/(8\pi a_s \ell^2)$~\citep{Dalfovo96}.
}
\label{t:units}
\end{table}

\subsection{Condensate wave function}

\begin{figure}[htbp] 
   \centering
   \includegraphics[width=7.5cm]{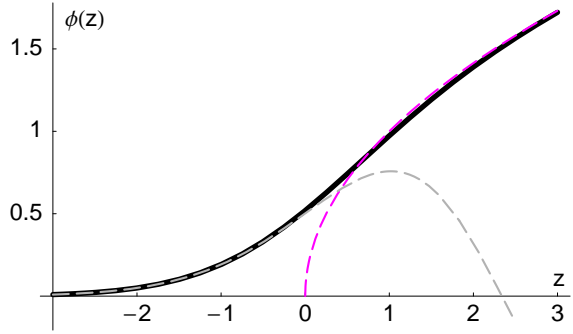} 
   \caption[]{Condensate wave function 
   {}(second Painlev\'e
   transcendent and solution
   to Eq.(\ref{eq:GPE})) and its asymptotic 
   behaviour (Eqs.(\ref{eq:dilute-asymptote}, \ref{eq:dense-asymptote}),
   dashed). We keep only the leading term in Eq.(\ref{eq:dense-asymptote}).}
   \label{fig:2nd-Painleve}
\end{figure}
The physically relevant
solution to Eq.(\ref{eq:GPE}) is known as the second Painlev\'e
transcendent and interpolates smoothly from an exponentially
decreasing (tunnelling) wave to 
the Thomas-Fermi solution obtained by neglecting the second 
derivative (Fig.\ref{fig:2nd-Painleve}).
Since one deals with a nonlinear equation, the amplitude of the
tunnelling solution (Airy function) is not arbitrary, and it 
has been shown that~\citep{Ablowitz77,Hastings80,Dalfovo96,Lundh97}
\begin{equation}
z \to -\infty: \qquad
\phi( z ) \to \sqrt{ 2 } \, {\rm Ai}( - z )
\label{eq:dilute-asymptote}
\end{equation}
On the dense side, \citet{Lundh97} and \citet{Margetis00} have 
improved the Thomas-Fermi solution into the expansion
\begin{equation}
z \to + \infty: \qquad
\phi( z ) \to
\sqrt{ z } \Big( 1 - 
\frac{ c_1 }{ z^3 }
- 
\frac{ c_2 }{ z^6 }
- \ldots
\Big)
\label{eq:dense-asymptote}
\end{equation}
with coefficients $c_1 = 1/8$, $c_2 = 73/128$, \ldots 
The condensate density appears in the \BdG{} Eqs.(\ref{eq:BdG}) 
for $u$ and $v$, for example via the Hartree-Fock potential
\begin{equation}
V_{\rm HF}( z ) = - z + 2 |\phi|^2
\to \left\{
\begin{array}{ll}
- z & \mbox{for\ }z \ll -1
\\
+ z & \mbox{for\ }z \gg +1
\end{array}
\right.
\label{eq:HF-wedge}
\end{equation}
which is a wedge-shaped trap
(Fig.\ref{fig:pot-wells}(\emph{left}), thin solid line). 
Numerically, we find a polynomial approximation near its bottom
(error $< 0.01$)
\begin{eqnarray}
V_{\rm HF}( z ) &\approx&
v_0 + v_2 ( z - z_0)^2 
+ v_3 (z - z_0)^3
\label{eq:HF-parabola}
\\
&& \mbox{for\ }
|z - z_0| \le 1
\nonumber
\end{eqnarray}
with the minimum located at $v_0 \approx 0.53$ and
$z_0 \approx 0.13$ and parameters $v_2 \approx 0.47$,
$v_3 \approx 0.041$. It will turn out, however, that the Hartree-Fock
well is irrelevant for the Bogoliubov solutions -- the only
message to keep is the characteristic energy scale $E = {\cal O}( 1 )$.

\subsection{Boundary conditions for Bogoliubov solutions}

The Bogoliubov
modes feature an intermediate zone $- E \lesssim z \lesssim E$
where the excitation changes its character from `single-particle' to
`collective'. Outside this zone, the asymptotic behaviour is as follows.

On the dilute side, the condensate $\phi( z )$ in Eqs.(\ref{eq:BdG})
vanishes, and the mode functions $u$ and $v$ decouple.
The linear branch of the Hartree-Fock potential 
$V_{\rm HF}( z ) \approx - z$
is a good approximation. We thus have a turning point $z_E$ 
for $u( z )$ at $z_E \approx - E$.
The mode $v( z )$ is already in the tunneling regime for $z \lesssim -1$
because of the opposite sign of the energy eigenvalue 
in Eq.(\ref{eq:BdG}).

\begin{figure}[htbp] 
   \centering
   \includegraphics[width=7.cm]{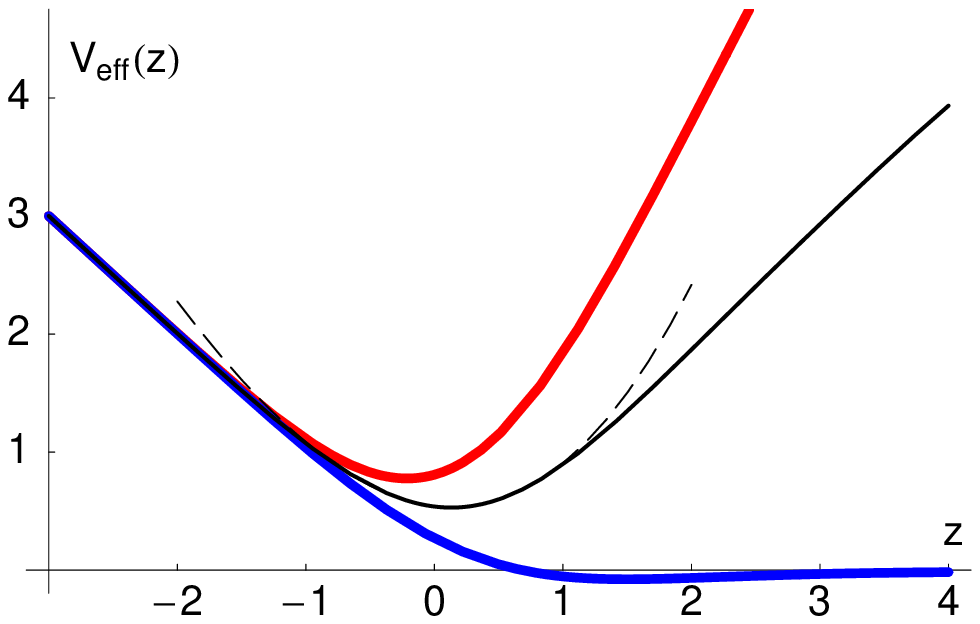} 
   \includegraphics[width=7.cm]{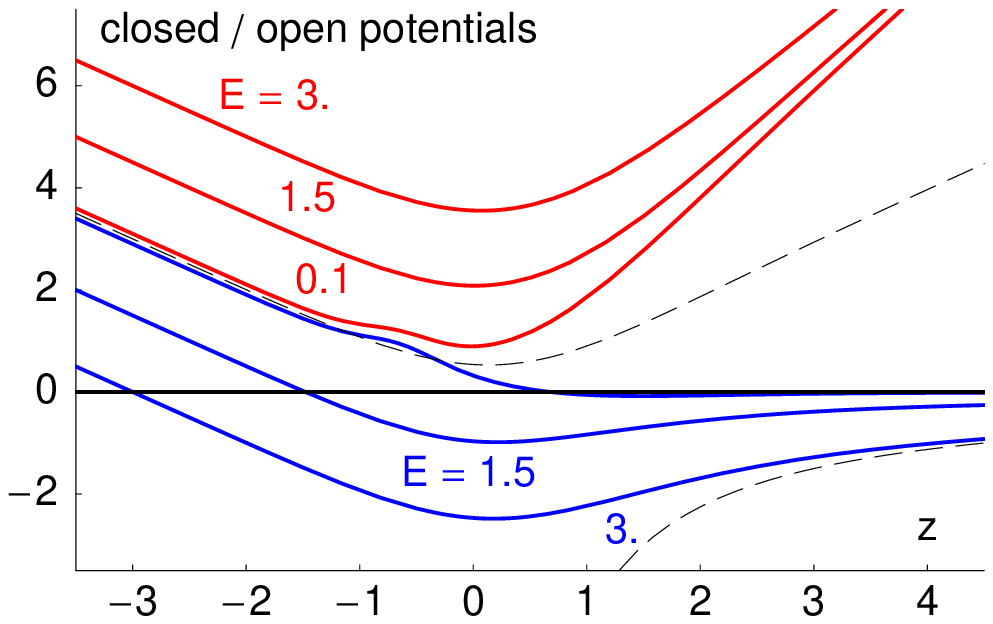} 
   \caption[]{(\emph{left}) Illustration of the `Hartree-Fock potential'
   (Eq.(\ref{eq:HF-wedge}), thin solid line, middle) and its variants
   that appear in the equations~(\ref{eq:phase-density-modes}) 
   for `phase' (blue, bottom) and `density' (red, top) modes.
   Dashed: parabolic approximation~(\ref{eq:HF-parabola})
   to the Hartree-Fock potential.
   \\
   (\emph{right}) Potentials in the adiabatic approximation for three
   energies. Upper (red) curves: 
   `density mode' $\tilde \kappa^2( z )$,
   lower (blue) curves: `phase mode' $- \tilde k^2( z )$ 
   (see Eqs.(\ref{eq:kappa-potl}, \ref{eq:k-potl})). As the energy $E$ 
   or the condensate mean-field potential $\phi^2( z )$ 
   increases, the potentials are pushed apart. The physical
   mode functions correspond, in this representation, to solutions 
   at zero energy (thick black line).  
   The bump around $-1 \lesssim z \lesssim 0$
   at low energies is due to the geometric potential,
   see discussion after Eqs.(\ref{eq:k-potl}, \ref{eq:kappa-potl})
   below.
   Upper dashed line: Hartree-Fock potential (Eq.(\ref{eq:HF-wedge}), 
   see left panel),
   lower dashed line: Coulomb-like asymptote of 
   Eq.(\ref{eq:retrieve-Coulomb}).
   }
   \label{fig:pot-wells}
\end{figure}
In the dense region where the condensate dominates, also the coupling
$\sim \phi( z )^2$ between $u$ and $v$ becomes large. It is convenient
to switch to the `density--phase' representation
$f = (u + v)/\sqrt{2}$ and $g = (u - v)/\sqrt{2}$ where the equations
become
\begin{eqnarray}
- \frac{ {\rm d}^2 f }{ {\rm d}z^2 } + 
( 3 \phi^2 - z ) f &=& E g
\nonumber\\
- \frac{ {\rm d}^2 g }{ {\rm d}z^2 } +
( \phi^2 - z ) g &=& E f
	\label{eq:phase-density-modes}
\end{eqnarray}
The `potentials' that appear here are plotted in red and blue in
Fig.\ref{fig:pot-wells}(\emph{left}).
The `density mode' $f$ corresponds to a well (upper red) whose spectrum starts
above zero energy (the minimum value of $3\phi^2 - z$ is 
$\approx 0.78$ at $z \approx - 0.21$). It is `enslaved' to the
`phase mode' $g$ that appears as a source term Eq.(\ref{eq:phase-density-modes}), 
first line.
The potential for the phase mode is a smooth barrier
that crosses zero at $z \approx 0.8$ and vanishes for $z \gg 1$
(Fig.\ref{fig:pot-wells}(left), lower blue).
To take into account the density-phase coupling
proportional to $E$, we perform the adiabatic elimination 
$f \approx E g / ( 2 z )$, using the Thomas-Fermi
asymptote $3 \phi^2 - z \approx 2 z$ and neglecting the second 
derivative. This gives deep in the condensate the equation 
for the phase mode 
\begin{equation}
z \to \infty: \quad
- \frac{ {\rm d}^2 g }{ {\rm d}z^2 } 
- \frac{ E^2 }{ 2 z } g \approx 0
\label{eq:retrieve-Coulomb}
\end{equation}
This one-dimensional Coulomb problem has exact solutions that
are discussed in 
Sec.\ref{s:Bessel-Coulomb} below. To state the boundary conditions,
a simpler semiclassical (WKB)
treatment will
suffice. From Eq.(\ref{eq:retrieve-Coulomb}),
identify the local wavenumber $k( z ) = E / \sqrt{ 2 z }$ 
and calculate the action integral: one gets two
independent solutions from the real and imaginary parts of
\begin{equation}
z \to \infty: \quad
g( z ) \sim
\frac{ (2 z)^{1/4} }{ \sqrt{ E } }
\exp( {\rm i} E \sqrt{ 2 z } )
	\label{eq:complex-f}
\end{equation}
Since $f( z )$ is smaller by a factor $E/(2z)$, this expression will 
dominate
the behaviour of both $u(z)$ and $v(z)$ deep in the condensate.
We call this asymptote the `local density approximation' because the
WKB treatment assumes that the condensate density $\phi^2( z ) 
\approx z$ varies slowly enough. In terms of the wave number, we require
$|{\rm d}k / {\rm d} z| \ll k^2$ or $E \sqrt{8 z} \gg 1$. This condition
illustrates that the border region $z \sim 0$ and the low-energy range
$E \ll 1$ are actually challenging and require techniques beyond the
WKB approximation. For a discussion of this point, see~\citet{Diallo-phase}.

To summarise, the physically relevant boundary conditions are 
\begin{itemize}
\item[(i)] \emph{dilute domain} through the 
turning point $z \sim -E$, but away from the condensate border
\begin{equation}
z \ll -1: \qquad
\begin{array}{rcl}
u( z ) &=& \alpha\, {\rm Ai}( - E - z )
\nonumber\\
v( z ) &=& \beta\, {\rm Ai}( E - z )
\end{array}
\label{eq:dilute-Airy}
\end{equation}
This covers the tunnelling region where both Airy
functions become exponentially small.
At large energies, the solutions are such that $v( z )$ 
is much smaller than $u( z )$.

\item[(ii)] \emph{local density approximation} in the dense (condensate)
region
\begin{eqnarray}
&&
z \gg E, 1/E^2: \qquad
\nonumber
	\\
&& u( z ) = 
\frac{ (2 z)^{1/4} }{ \sqrt{ 2 E } } 
\cos( E \sqrt{ 2 z } - \pi / 4 + \delta )
\nonumber\\
&& v( z ) = 
-
\frac{ (2 z)^{1/4} }{ \sqrt{ 2 E } } 
\cos( E \sqrt{ 2 z } - \pi / 4 + \delta )
\label{eq:dense-with-phase}
\end{eqnarray}
We have considered here only the leading order terms 
proportional to the phase mode $g$ (Eq.(\ref{eq:complex-f})). 
The normalisation is such that the solutions $\delta = 0, \pi /2$
have the same amplitude and unit Wronskian, see \ref{a:Wronskian} 
for details.
\end{itemize}
The phase shift $\delta$ in the bulk asymptote~(\ref{eq:dense-with-phase})
depends on the relative weight between real and
imaginary parts of the complex solutions~(\ref{eq:complex-f}). 
The reference $-\pi / 4$ is explained in Sec.\ref{s:Bessel-Coulomb}.
We emphasise that this phase shift $\delta = \delta( E )$ 
`carries' information about the behaviour 
near the condensate border into the bulk. For a matching of 
the \BdG{} solutions near the turning point with bulk solutions
using boundary layer techniques, see for example~\citet{Fetter98a}.

\section{Phase and density modes}
\label{s:uv-modes}

The coupled \BdG{} equations contain
unphysical solutions that grow for $z \to \pm\infty$
and that typically contaminate numerical trials 
when the \BdG{} equations are straightforwardly integrated. 
This can be seen from the second line of
Eqs.(\ref{eq:phase-density-modes}) whose homogeneous solutions
are `under the barrier' and grow exponentially.
We have developed instead a semi-analytical scheme where the unstable
modes are eliminated. The idea is to perform a rotation in the $uv$-plane 
that diagonalises the coupling.

\subsection{Adiabatic transformation}

We make the following \emph{Ansatz} for a rotated set of amplitudes
\begin{equation}
\Bigg(
\begin{array}{c}
u \\ v
\end{array}
\Bigg)
=
\Bigg(
\begin{array}{cc}
\cos\theta/2 & \sin\theta/2
\\ 
-\sin\theta/2 & \cos\theta/2
\end{array}
\Bigg)
\Bigg(
\begin{array}{c}
\tilde u \\ \tilde v
\end{array}
\Bigg)
	\label{eq:def-rotation}
\end{equation}
and find that the coupling between $\tilde u$ and $\tilde v$ 
(Eqs.(\ref{eq:BdG}))
is removed when
the rotation angle $\theta$ is chosen as
\begin{equation}
	\tan \theta( z ) = \frac{ \phi^2( z ) }{ E }
	\label{eq:def-theta}
\end{equation}
Note that in the dense region, we have $\theta \to \pi/2$ and
the amplitudes $\tilde u$, $\tilde v$ approach the phase and
density modes ($g$, $f$) introduced above Eq.(\ref{eq:phase-density-modes}).
We note that a hyperbolic rotation that preserves the Bogoliubov
norm $u^2 - v^2$ can also be used, but leads only to minor changes 
in notation.

The equations for $\tilde u$ and $\tilde v$
do not decouple completely because 
the rotation angle $\theta$ depends on position.
By working out the second derivative of Eq.(\ref{eq:def-rotation}),
we get
\begin{eqnarray}
- \frac{ {\rm d}^2 \tilde u }{ {\rm d}z^2 } - 
\tilde k^2 \tilde u &=&
\Lop \tilde v
	\label{eq:BdG-adiabatic-u}
\\
- \frac{ {\rm d}^2 \tilde v }{ {\rm d}z^2 } + 
\tilde \kappa^2 \tilde v &=&
- \Lop \tilde u
	\label{eq:BdG-adiabatic}
\end{eqnarray}
where the coupling 
involves derivatives of the condensate density via the differential
operator
\begin{equation}
\Lop = {\textstyle\frac12} \theta'' 
+ \theta' \frac{ {\rm d} }{ {\rm d}z } 
	\label{eq:L-op}
\end{equation}
The `adiabatic potentials'
$- \tilde k^2$ and $\tilde\kappa^2$ 
in Eqs.(\ref{eq:BdG-adiabatic-u}, \ref{eq:BdG-adiabatic})
are recognised as the 
generalisations of the phase and density potentials
$\phi^2 - z$ and $3 \phi^2 - z$ of Eqs.(\ref{eq:phase-density-modes}).
They are plotted in Fig.\ref{fig:pot-wells}(\emph{right}) and take the form
\begin{eqnarray}
- \tilde k^2 &=& 
- z + 2 \phi^2
- \sqrt{ E^2 + \phi^4 } 
+ \big( {\textstyle\frac12} \theta' \big){}^2
	\label{eq:k-potl}
\\
\tilde\kappa^2 &=& 
- z + 2 \phi^2
+ \sqrt{ E^2 + \phi^4 } 
+ \big( {\textstyle\frac12} \theta' \big){}^2
	\label{eq:kappa-potl}
\end{eqnarray}
We can understand the additional term $(\frac12 \theta')^2$ 
in Eqs.(\ref{eq:k-potl}, \ref{eq:kappa-potl})
as a `geometric potential',
by analogy to the geometric phase for a spin that is adiabatically
transported in a slowly varying field~\citep{Berry84,WilczekBook}. 
Since we deal with a second-order differential equation, the
structure is slightly different from the conventional geometric phase: 
one might also call `geometric' the off-diagonal 
operator $\Lop$ in Eq.(\ref{eq:L-op}).
\begin{figure}[htbp] 
   \centering
   \includegraphics[width=7.5cm]{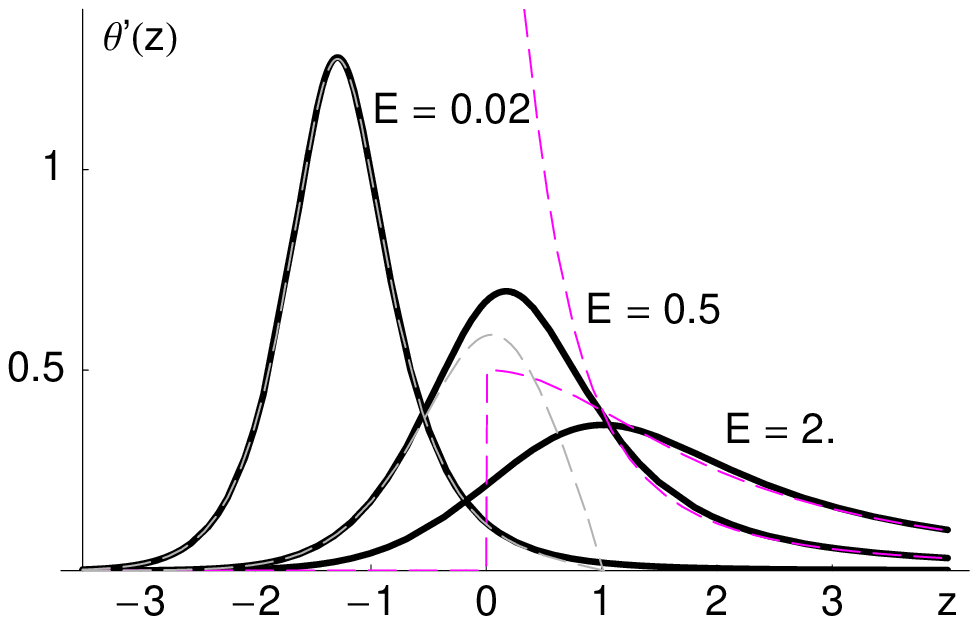} 
   \caption[]{Illustration of 
   
   the non-adiabatic coupling $\theta'( z )$ (see Eq.(\ref{eq:def-theta}))
   for selected energies.
   Dashed: based on Airy and Thomas-Fermi approximations to
   the condensate density, Eqs.(\ref{eq:dilute-asymptote}, 
   \ref{eq:dense-asymptote}).}
   \label{fig:illu-theta}
\end{figure}
This operator,
involving the derivatives $\theta' = 
{\rm d}\theta / {\rm d}z$ and $\theta''$, is called `non-adiabatic
coupling' in the following. It peaks roughly where the
mean-field potential $\phi^2(z)$ crosses the mode energy $E$, as
illustrated in Fig.\ref{fig:illu-theta}. The Thomas-Fermi approximation
$\theta( z ) \approx \arctan( z / E )$ provides a simple overview in 
the dense region, for example
(magenta dashed in Fig.\ref{fig:illu-theta}):
\begin{equation}
z \gg 1: \qquad
\theta'( z ) \approx \frac{ E }{ z^2 + E^2 }
\label{eq:TF-for-thetap}
\end{equation}
The non-adiabatic couplings are thus confined to the 
`condensate border' $z \lesssim E$ and become weak as the energy grows. 
Conversely, for $E \to 0$, the
maximum of $\theta'( z )$ shifts into the dilute region with a scaling
in position (height) roughly proportional to 
$-(\log 1/E)^{2/3}$ ($(\log 1/E)^{1/3}$), respectively 
as can be checked from
the tunnelling asymptotics of the Airy function 
{}(dashed gray in Fig.\ref{fig:illu-theta}).

In the following, we proceed by adopting first the adiabatic
approximation where the off-diagonal terms proportional to $\Lop$
are neglected (Sec.\ref{s:Bessel-Coulomb}). 
Non-adiabatic corrections are discussed in 
Sec.\ref{s:trapping-states}, in particular the role they play
for the `density mode' $\tilde v$.

\subsection{Phase modes in open potential}
\label{s:Bessel-Coulomb}

\begin{figure}[htbp] 
   \centering
   \includegraphics[width=7.5cm]{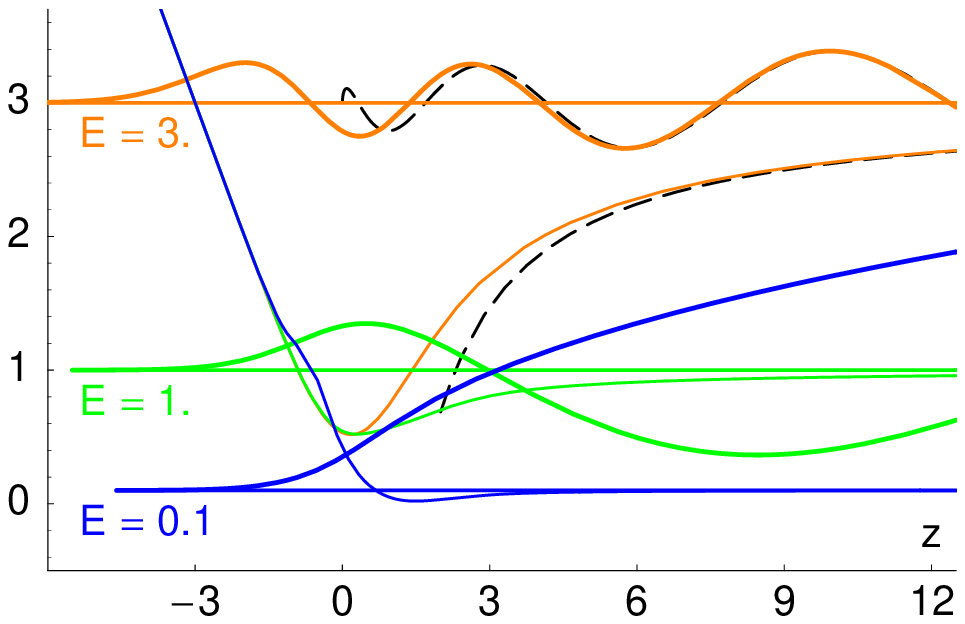} 
   \includegraphics[width=7.5cm]{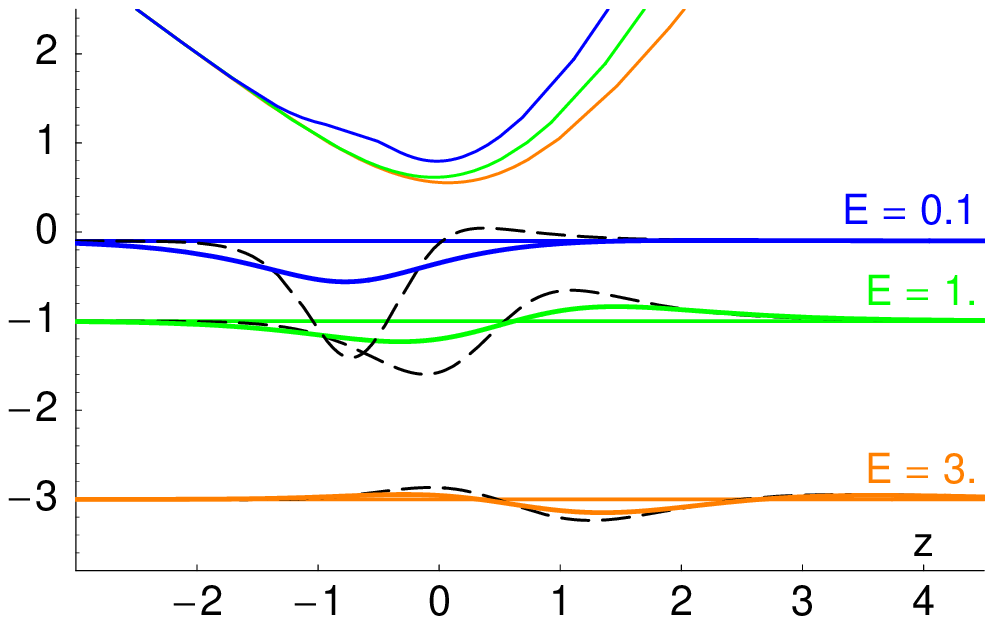} 
   \caption[]{(\emph{left}) Bogoliubov phase mode $\tilde u$ in the
   open channel for different energies, using the adiabatic approximation. 
   For the ease of comparison with Fig.\ref{fig:pot-wells},
   we have plotted shifted potentials $E - \tilde k^2( z )$. The wave
   functions $\tilde u( z )$ are multiplied by $1/\sqrt{10}$.
   The bump in the potential around $z = -1$
   at low energy is due to the geometric correction
   $[\frac12\theta'( z )]^2$.
   Black dashed: Coulomb tail of the potential $-\tilde k^2( z )$, 
   as given in Eq.(\ref{eq:Coulomb-tail}), and corresponding regular
   solution (Eq.(\ref{eq:Bessel-Coulomb-solutions})).
   \\
   (\emph{right}) Closed-channel or density modes $\tilde v$,
   calculated perturbatively from the adiabatic approximation
   $\tilde u( z )$. We have shifted the potentials to
   $\tilde \kappa^2( z ) - E$ so that the
   wave functions appear at the energy $-E$, as expected from
   Eq.(\ref{eq:BdG}); they have been multiplied by $\sqrt{10}$
   for better visibility.
   Dashed lines: simple adiabatic elimination
   $\tilde v \approx - L \tilde u / \kappa^2( z )$.
   }
   \label{fig:u-waves}
\end{figure}

\noindent
In the adiabatic approximation (subscript `${\rm ad}$'), 
the equation for $\tilde u$ can be written in the form
\begin{equation}
-
\frac{ {\rm d}^2 \tilde u_{\rm ad} }{ {\rm d}z^2 }
+ \left( E -
\tilde k^2( z ) \right) \tilde u_{\rm ad} 
= E \tilde u_{\rm ad} 
	\label{eq:open-Schroedinger}
\end{equation}
where the potential $E -\tilde k^2( z )$ is given by 
Eq.(\ref{eq:k-potl}) for all $z$ (Fig.\ref{fig:u-waves}(\emph{left})). 
At low energies, it is similar to the 
lower (blue) curve in Fig.\ref{fig:pot-wells}.
We call it an `open channel' because we have $E - \tilde k^2( z ) \le E$
as $z \to \infty$ so that $\tilde u$ is an extended wave right
at the continuum threshold, with a turning point near $z = - E$,
as shown in the Figure.
\ref{a:numerics} provides some details on the numerical calculation
of these solutions.

Deep in the condensate, we find (black dashed line in 
Fig.\ref{fig:u-waves}(\emph{left}))
\begin{eqnarray}
	&&
	 \phi^2( z ) \gg E:
\nonumber
	\\
&&
	 - \tilde k^2( z ) \approx 
	 - \frac{ E^2 }{ 2 z } 
	 - \frac{ 1 }{ 4 z^2 }
	 + {\cal O}( z^{-5}, E^2 z^{-4}, E^4 z^{-3} )
	\label{eq:Coulomb-tail}
\end{eqnarray}
where the first term recovers the approximation~(\ref{eq:retrieve-Coulomb}).
The `centripetal term' $\sim 1 / z^2$ arises from the first correction
beyond the Thomas-Fermi approximation (the coefficient 
$c_1 = 1/8$ in Eq.(\ref{eq:dense-asymptote})).
The higher-order corrections arise from the next-to-leading order 
expansion of the root $\sqrt{ E^2 + \phi^4( z ) }$
and from the geometric potential $[\frac12 \theta'(z)]^2$.
The Schr\"odinger equation for $\tilde u_{\rm ad}( z )$ therefore 
matches asymptotically with a modified Coulomb problem:
\begin{eqnarray}
&&
- \frac{ {\rm d}^2 \psi }{ {\rm d}z^2 } 
	 + V_C( z ) \psi
	 = 0
	\label{eq:modified-Coulomb}
\\
&&	
V_C( z ) = 
	 - \frac{ E^2 }{ 2 z } 
	 - \frac{ 1 }{ 4 z^2 }
\nonumber
\end{eqnarray}
an equation that replaces Eq.(\ref{eq:retrieve-Coulomb}) obtained
above with a simpler argument.
The required solutions are located just at the dissociation threshold
of the Coulomb potential;
they are known analytically and are 
linear combinations of Bessel functions~\citep{Abramowitz} 
{}(black dashed in Fig.\ref{fig:u-waves}(\emph{left}))
\begin{eqnarray}
	j( z ) &=&
	\sqrt{ \pi z }\,
	J_0( E \sqrt{ 2 z } )
\nonumber\\
	y( z ) &=&
	\sqrt{ \pi z }\,
	Y_0( E \sqrt{ 2 z } )
	\label{eq:Bessel-Coulomb-solutions}
\end{eqnarray}
The argument $E \sqrt{ 2 z }$ of the Bessel 
functions is familiar from the phase of the WKB 
solutions in Eq.(\ref{eq:complex-f}).
We have chosen a normalisation such that both Bessel-Coulomb solutions 
have the same amplitude deep in the condensate and their
Wronskian is equal to unity \citep{Abramowitz}
\begin{equation}
W[j, y] = j y' - y j' 
= j \frac{ {\rm d}y }{ {\rm d}z } 
- y \frac{ {\rm d}j }{ {\rm d}z }
\label{eq:def-Wronskian}
\end{equation}
Deep in the condensate, the Bogoliubov mode can therefore be
represented in the form
\begin{eqnarray}
&& z \to \infty: \quad
\nonumber\\
&&
\tilde u_{\rm ad}( z ) \approx
{\cal A} \Big(
	j( z ) \cos \delta_{\rm ad}
	- 
	y( z ) \sin \delta_{\rm ad} 
	\Big)
\label{eq:def-phase-shift}
\end{eqnarray}
where ${\cal A}$ is a normalisation.
This formula \emph{defines} the phase shift $\delta_{\rm ad} 
= \delta_{\rm ad}( E )$ 
of the Bogoliubov mode: the reference case $\delta_{\rm ad} = 0$ corresponds 
to the Coulomb wave $j( z )$ which is regular when extrapolated back 
to the condensate border (at $z = 0$ in the Thomas-Fermi approximation).
According to the 
asymptotic series of $J_0$, $Y_0$, the Bogoliubov mode function
will match the behaviour deep in the condensate
we required in Eq.(\ref{eq:dense-with-phase}) above:
\begin{eqnarray}
&& E \sqrt{ 2 z } \gg 1: \qquad
\nonumber\\
&&
\tilde u_{\rm ad}(z) \approx
\frac{ (2 z)^{1/4} }{ \sqrt{ E } } \cos( E \sqrt{2 z} - \pi/4 
	+ \delta_{\rm ad})
\label{eq:asymptotics-phase-shift}
\end{eqnarray}
provided we choose the normalisation factor ${\cal A} = 1$
in Eq.(\ref{eq:def-phase-shift})
{}(see~\ref{a:Wronskian}, Eq.(\ref{eq:completeness-uv})).
We recall that $\cos \theta/2, \sin \theta/2 \to 1/\sqrt{2}$ in this limit
and that $\tilde v( z ) = 0$ in the adiabatic approximation.

Note that the 
`centripetal potential' $- 1/(4z^2)$ 
that arises from the first `post-Thomas-Fermi' correction
$\phi^2( z ) \approx z - 1/(4z^2)$ 
in Eq.(\ref{eq:dense-asymptote})
is significant in this context. Dropping it
from Eq.(\ref{eq:modified-Coulomb}), the analytical solutions
would involve first-order Bessel functions $J_1$, $Y_1$ which are
phase-shifted by $\pi/2$ relative to their zeroth-order counterparts.
This could have been expected from the long-range character of the
centripetal potential, on the one hand. On the other, it is interesting 
to realise that one needs the $J_0$ function in 
Eq.(\ref{eq:Bessel-Coulomb-solutions}) 
to recover the correct behaviour of the Bogoliubov modes 
at low energies, as required by the U(1) global phase symmetry of
the mean field theory. We discuss the low-energy limit
in more detail in Sec.\ref{s:low-energies}.

\subsection{Non-adiabatic coupling and density modes in closed potential}
\label{s:trapping-states}

We now take into account the off-diagonal (coupling) terms in the 
\BdG{} equations and solve Eq.(\ref{eq:BdG-adiabatic})
for the density mode
\begin{equation}
-
\frac{ {\rm d}^2 \tilde v_{\rm ad} }{ {\rm d}z^2 }
+
\tilde \kappa^2( z ) \tilde v_{\rm ad} 
= - \Lop \tilde u_{\rm ad}( z )
	\label{eq:closed-Schroedinger-ad}
\end{equation}
This mode influences
significantly the scattering phase shift $\delta$, as we discuss
in a companion paper~\citep{Diallo-phase}. In addition, it also
contributes to the spectrum of density fluctuations (dynamical structure 
factor), as illustrated in Sec.\ref{s:density-fluctuations} below.

\begin{figure}[htbp] 
\begin{tabular}{c@{\hspace*{-3mm}}c}
\begin{tabular}{c}
	\includegraphics[width=7.5cm]{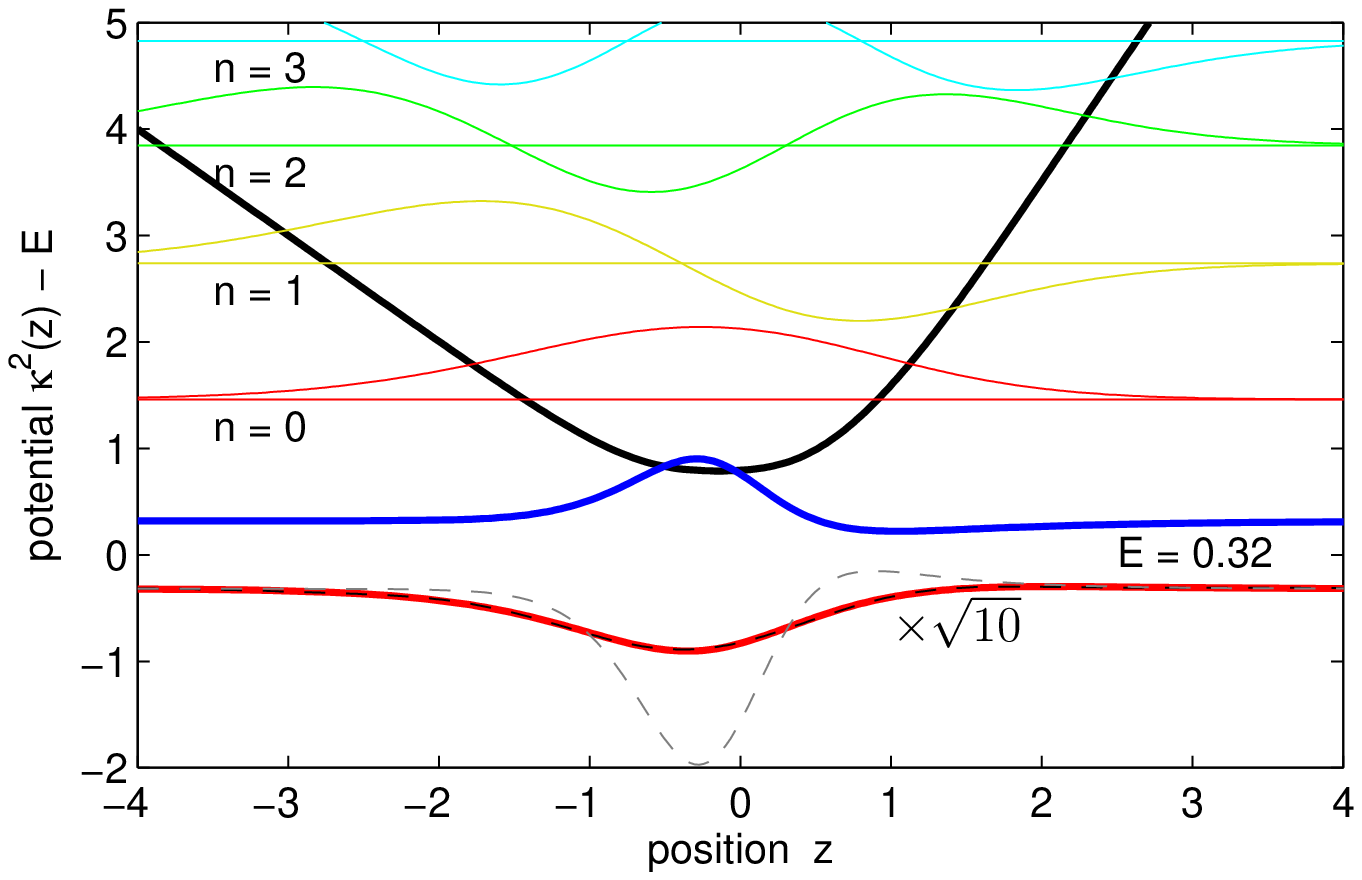} 
	\\[-05mm]
	\includegraphics[width=7.5cm]{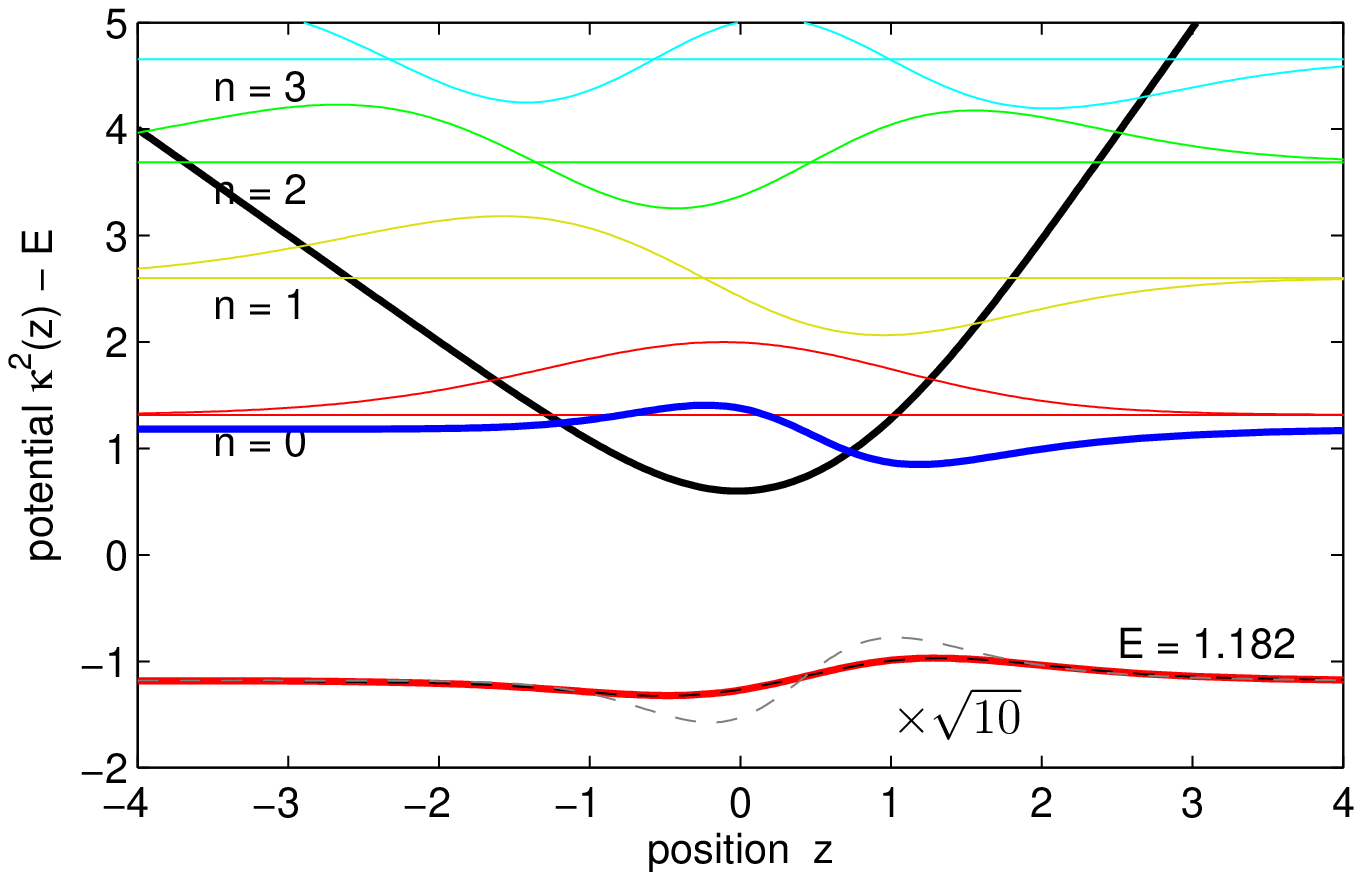} 
\end{tabular}
&  
\raisebox{-30mm}{\includegraphics[width=8.5cm]{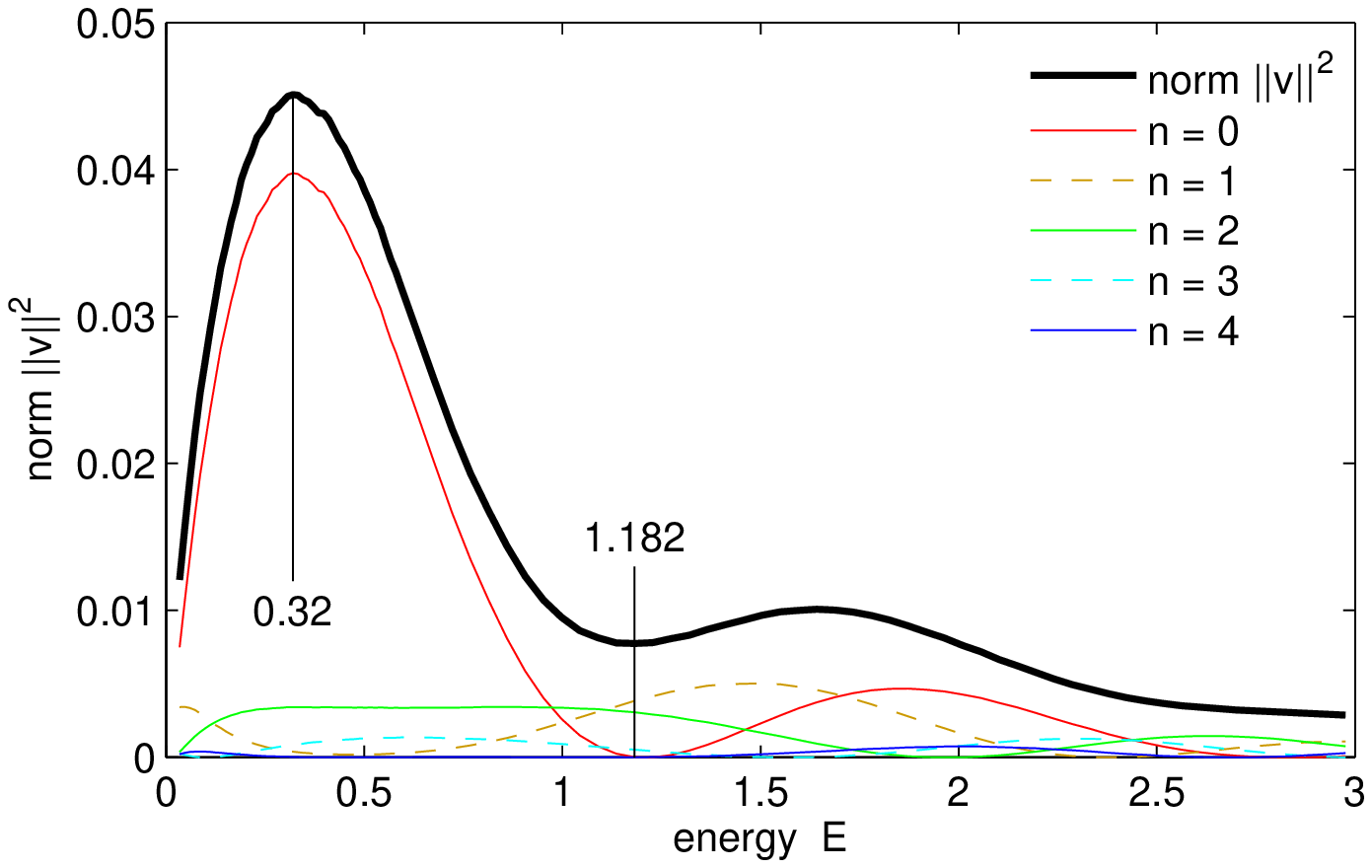}}
\end{tabular}
   \caption[]{(\emph{left column}) 
   Solutions to the eigenvalue problem in the
   closed channel $\tilde\kappa^2( z )$ (thick black), shifted by $-E$
   ($n = 0 \ldots 3$).
   We also plot the source term $\Lop \tilde u$ (thick blue) at baseline $+E$.
   
   The density mode $\tilde v_{\rm ad}( z )$ is shown
   at baseline $-E$, solved from the inhomogeneous Schr\"odinger
   Eq.(\ref{eq:closed-Schroedinger-ad}) in the closed channel.
   Thick red = direct numerical inversion of the wave operator; 
   overlapping with dashed black = expansion into
   the lowest twelve trapped modes according to 
   Eqs.(\ref{eq:expand-tilde-v}, \ref{eq:solution-trapped-amplitudes});
   dashed gray: local approximation, neglecting the second derivative.   
   \\
   (\emph{right}) Squared norm $\Vert \tilde v \Vert^2$ 
   (Eq.(\ref{eq:def-v-norm})) of the density mode
   as a function of energy (upper thick line). 
   This
   is compared to the squares $|b_n|^2$ 
   (Eq.(\ref{eq:solution-trapped-amplitudes})) 
   of its expansion into low-lying
   trapped modes (thin lines, odd modes dashed). 
   At the marked energies, the overlap to the ground mode is
   maximal and minimal, as shown in the left column.
   }
   \label{fig:trapped-modes}
\end{figure}

The Schr\"odinger operator on the left-hand side of 
Eq.(\ref{eq:closed-Schroedinger-ad}) corresponds to
a wedge-shaped potential well whose minimum is above zero
{}(Fig.\ref{fig:u-waves}(\emph{right})). If the right-hand
side is neglected, 
we therefore do not have any physically acceptable solution
that remains finite for $z \to \pm \infty$. 
Numerically, the inhomogeneous equation can be solved straightforwardly 
by representing the second derivative with a finite difference scheme
and solving the corresponding
sparse linear system. The results are shown in 
Fig.\ref{fig:u-waves}(\emph{right}) and
Fig.\ref{fig:trapped-modes}(\emph{left}).
As expected, the density mode is
localised in the border region and has an
amplitude much smaller than the phase mode. The `local approximation'
$\tilde v = - L \tilde u / \kappa^2(z)$ (gray dashed line) captures
well its tails, but not the reduced amplitude 
of the oscillatory features (where the second derivative is obviously
significant). 

Some insight into the inhomogeneous Schr\"odinger 
Eq.(\ref{eq:closed-Schroedinger-ad}) may be gathered 
by considering first the eigenvalue problem
\begin{equation}
-
\frac{ {\rm d}^2 v_n }{ {\rm d}z^2 }
+
\tilde \kappa^2( z ) v_n 
= \epsilon_n v_n( z )
	\label{eq:closed-eigenstates}
\end{equation}
for the potential well provided by $\tilde \kappa^2( z )$.
The eigenmodes $v_n$
provide a convenient basis to expand $\tilde v$:
\begin{equation}
\tilde v( z ) = \sum_{n} b_n v_n( z )
	\label{eq:expand-tilde-v}
\end{equation}
The coefficients $b_n$ are found by projecting 
Eq.(\ref{eq:closed-Schroedinger-ad}) onto $v_n$, using the natural
scalar product
\begin{equation}
( v_n | \tilde v ) = \int\!{\rm d}z\, v_n( z ) \tilde v( z )
	\label{eq:def-L2-norm}
\end{equation}
and choosing the normalisation $(v_n | v_m) = \delta_{nm}$.
We get after two partial integrations
\begin{equation}
b_n = - \frac{ (v_n | \Lop \tilde u_{\rm ad}) }{ \epsilon_n }
	\label{eq:solution-trapped-amplitudes}
\end{equation}
One key property here is that the source term in
Eq.(\ref{eq:closed-Schroedinger-ad}), 
$\Lop \tilde u_{\rm ad}( z )$,
is actually a localised function (Fig.\ref{fig:trapped-modes}(\emph{left}),
thick blue) 
because the differential operator $\Lop$ involves the derivatives
$\theta'( z )$ and $\theta''( z )$ that tend to zero
as $z \gg E$ (see Eq.(\ref{eq:TF-for-thetap})).
The matrix elements in Eq.(\ref{eq:solution-trapped-amplitudes}) 
are therefore given by convergent integrals.
The `trapped modes' $v_n$ are illustrated in 
Fig.\ref{fig:trapped-modes}(\emph{left}).
The closed-channel potential $\tilde \kappa^2( z )$
is harmonic only in a narrow range around its minimum.
Hence, the spectrum is non-equidistant due to the linear asymptotes
away from the minimum. We derive in~\ref{a:trap-spectrum} the
asymptotics $\epsilon_n \sim E + [\pi (n + \frac12)]^{2/3}$.

\paragraph{Pseudo-Feshbach resonance.}
The results for the trapped density mode
are summarised in Fig.\ref{fig:trapped-modes}(\emph{right})
where we plot the norm of the density mode $\tilde v_{\rm ad}$
{}(defined as in Eq.(\ref{eq:def-v-norm}) below) vs.\ the energy $E$.
To interpret the oscillating features, we suggest an 
analogy to the so-called
Feshbach resonances in atomic and molecular scattering. The physics
is essentially the same: due to non-adiabaticity, different potential
surfaces are coupled. The colliding system may thus split up
and follow different
paths which eventually interfere in the output (St\"uckelberg 
oscillations). A particularly strong
effect occurs when a localised eigenstate in a closed channel becomes
degenerate with the incoming wave in an open channel. 
In ultracold collisions, this mechanism operates when a differential Zeeman 
shift brings coupled spin states into resonance; the result is
a divergence of the scattering length for specific values of
the magnetic field (Feshbach resonance).

In our problem, we also have two potentials, open ($-\tilde k^2$) 
and closed ($\tilde \kappa^2$). But there is no tuning parameter 
available to bring the initial wave into resonance
with the closed-channel eigenvalues: the denominators $\epsilon_n$
in the amplitudes $b_n$ (Eq.(\ref{eq:solution-trapped-amplitudes}))
never cross zero. 
One can even derive the stronger bound 
$\epsilon_n \ge E + v_0 + \sqrt{v_2} \approx E + 1.22$ 
from the ground state of the harmonic approximation to the Hartree-Fock 
potential contained in $\tilde \kappa^2( z )$ (see Eq.(\ref{eq:HF-parabola})).
There is, however, one possibility for a resonantly enhanced
density mode. It is not related to a matching of
energies, but of wave functions. Indeed, 
for the energy $E \approx 0.32$,
one observes a quite accurate matching in shape and position
between $L \tilde u_{\rm ad}$ and the ground state $v_0$
{}(Fig.\ref{fig:trapped-modes}(\emph{top left})).
This leads (as in a Franck--Condon argument)
to the strong peak in the norm of the density mode
\begin{equation}
\Vert \tilde v \Vert^2 = (\tilde v | \tilde v) 
= \int\!{\rm d}z\, \tilde v^2( z )
	\label{eq:def-v-norm}
\end{equation}
as can be seen in
Fig.\ref{fig:trapped-modes}(\emph{right}) 
where the probabilities $|b_n|^2$ are plotted 
as a function of energy $E$
and compared to the norm $\Vert \tilde v \Vert^2$.
A resonance with the first excited state
$v_1$ at $E \approx 1.5$ is visible because at a slightly lower
energy (Fig.\ref{fig:trapped-modes}(\emph{bottom left})), 
$\Lop \tilde u_{\rm ad}$ becomes orthogonal to the ground state $v_0$.
At this `anti-resonance', the derivative term in $\Lop$ is significant.

We have observed that
the shape of the closed-channel potential $\tilde \kappa^2( z )$
is relatively stable as the energy $E$ increases
{}(compare Fig.\ref{fig:trapped-modes}(left, top and bottom)). 
The overlap
therefore changes chiefly because the turning point and the
nodes of the open-channel solution $\tilde u$ shift with $E$,
as we saw in Fig.\ref{fig:u-waves}(\emph{left}).
The other reason is the shifting
and broadening of the non-adiabatic couplings $\theta'$, 
$\theta''$ that are involved in the operator $\Lop$
(recall Fig.\ref{fig:illu-theta}).

\subsection{Low-energy behaviour}
\label{s:low-energies}

It is well known that when the Bogoliubov spectrum is continuous,
it is gapless and that the
amplitudes $u$ and $v$ approach the shape of the condensate
in the limit $E \to 0$. This translates the Goldstone mode arising
from the global phase invariance (U(1) symmetry) of the Gross-Pitaevskii 
equation~(\ref{eq:GPE}). In this low-energy limit, the phase-density 
representation
of Eqs.(\ref{eq:phase-density-modes}) becomes exact. For the sake of
simplicity, we stay in the adiabatic basis~(\ref{eq:def-rotation})
and get in the leading order the \BdG{} equation for the phase mode
\begin{equation}
E \to 0: \qquad
- \frac{ {\rm d}^2 \tilde u }{ {\rm d}z^2 } + 
( \phi^2 - z ) \tilde u = 0
\label{eq:zero-E-phase}
\end{equation}
This is solved by the condensate $\phi( z )$ itself.
{}(For an illustration, see the $E = 0.1$ curve in 
Fig.\ref{fig:u-waves}(\emph{left}).)
In the same limit, there is only the trivial solution $\tilde v = 0$
for the density mode. 
We fix the normalisation by continuity with the low-energy
limit of the Bessel-Coulomb wave that is proportional to the
Thomas-Fermi condensate:
\begin{eqnarray}
&&
E^2 \lesssim E^2 z \ll 1: \qquad
\nonumber\\
&&
\tilde u( z ) = \sqrt{ \pi } \phi( z )
\approx j( z ) = \sqrt{ \pi z } ( 1 + {\cal O}( E^2 z ) )
\label{eq:low-E-tilde-u}
\end{eqnarray}
We indeed find that the phase shift $\delta( E )$ is very small
in this limit
so that the other Bessel-Coulomb wave $y( z )$ (see Eq.(\ref{eq:def-phase-shift})) 
has negligible weight at low energies \citep{Diallo-phase}. 
The spatial range $1 \lesssim z \ll 1/E^2$ where this behaviour
is relevant opens up wide for $E \to 0$.

\section{Applications}
\label{s:applications}

\subsection{Equilibrium correlations}

It is well known that the Bogoliubov--de Gennes modes provide 
a convenient expansion of the field operator
\begin{equation}
\psi( z ) = 
\phi( z ) 
+
\int\limits_{0}^{\infty}
\!\frac{ {\rm d}E }{ \sqrt{ \pi } }
\left\{ 
u_E( z ) a_E + v_E( z ) a_E^\dag
\right\}
\label{eq:expand-field}
\end{equation}
where the operators $a_E^\dag$ ($a_E$) create (annihilate)
an elementary excitation with energy $E$.
We have assumed a c-number
condensate (Bogoliubov shift) for simplicity and added 
the subscript $E$ to the mode functions for clarity.
Since the inhomogeneous potential is `open' on the dense side,
the energy spectrum is continuous.
{}(The integration measure ${\rm d}E / \sqrt{ \pi }$ arises from
the normalisation of the $u$, $v$, see~\ref{a:Wronskian}.)
In thermal equilibrium, we have $\langle a_E \rangle = 0$
and the Bose occupation number
\begin{equation}
\langle a_E^\dag a_{E'} \rangle 
= \bar N(E) \delta( E - E' )
= \frac{ \delta( E - E' ) }{ {\rm e}^{ E / T } - 1 }
\label{eq:thermal-expectation}
\end{equation}
because the expansion~(\ref{eq:expand-field}) provides 
a quadratic approximation of the
second-quantised field Hamiltonian. 
The elementary excitations contribute even
at zero temperature (`depletion') because of the operator $a_E^\dag$ that
appears in Eq.(\ref{eq:expand-field}). We focus in the following
on low temperatures and leave aside the problem of 
`quasi-condensation' and self-consistent mean-field theories
in low dimensions; see for 
example \citet{Andersen02c, AlKhawaja02b, Mora03}.

\subsection{Field correlation spectrum}
\label{s:field-fluctuations}

Matter-wave interference experiments are sensitive to the dynamic field 
correlation function
\begin{equation}
G( x, y, \tau ) = \langle
\psi^\dag( x, t ) \psi( y, t + \tau )
\rangle
\label{eq:def-field-corr}
\end{equation}
where the time dependence arises from
$a_E( t ) \sim {\rm e}^{ - {\rm i} E t }$ in the Heisenberg
picture. (Recall that we have set the zero of energy at the
chemical potential.)
Inserting Eq.(\ref{eq:expand-field}) and taking $x = y$, we get 
the well-known expression
\begin{eqnarray}
&& G(z, \tau ) = |\phi( z )|^2 
\nonumber\\
&& {} 
+ \int\limits_{0}^{\infty}\!\frac{ {\rm d}E }{ \pi }
\left\{
u_E^2( z ) \bar N( E ) \, {\rm e}^{ - {\rm i} E t }
+
v_E^2( z ) ( 1 + \bar N( E ) ) \, {\rm e}^{ {\rm i} E t }
\right\}
\label{eq:spectrum-field-corr}
\end{eqnarray}
\begin{figure}[htbp] 
\begin{tabular}{c@{\hspace*{-01mm}}c}
\begin{tabular}{c}
   \includegraphics*[width=6.5cm]{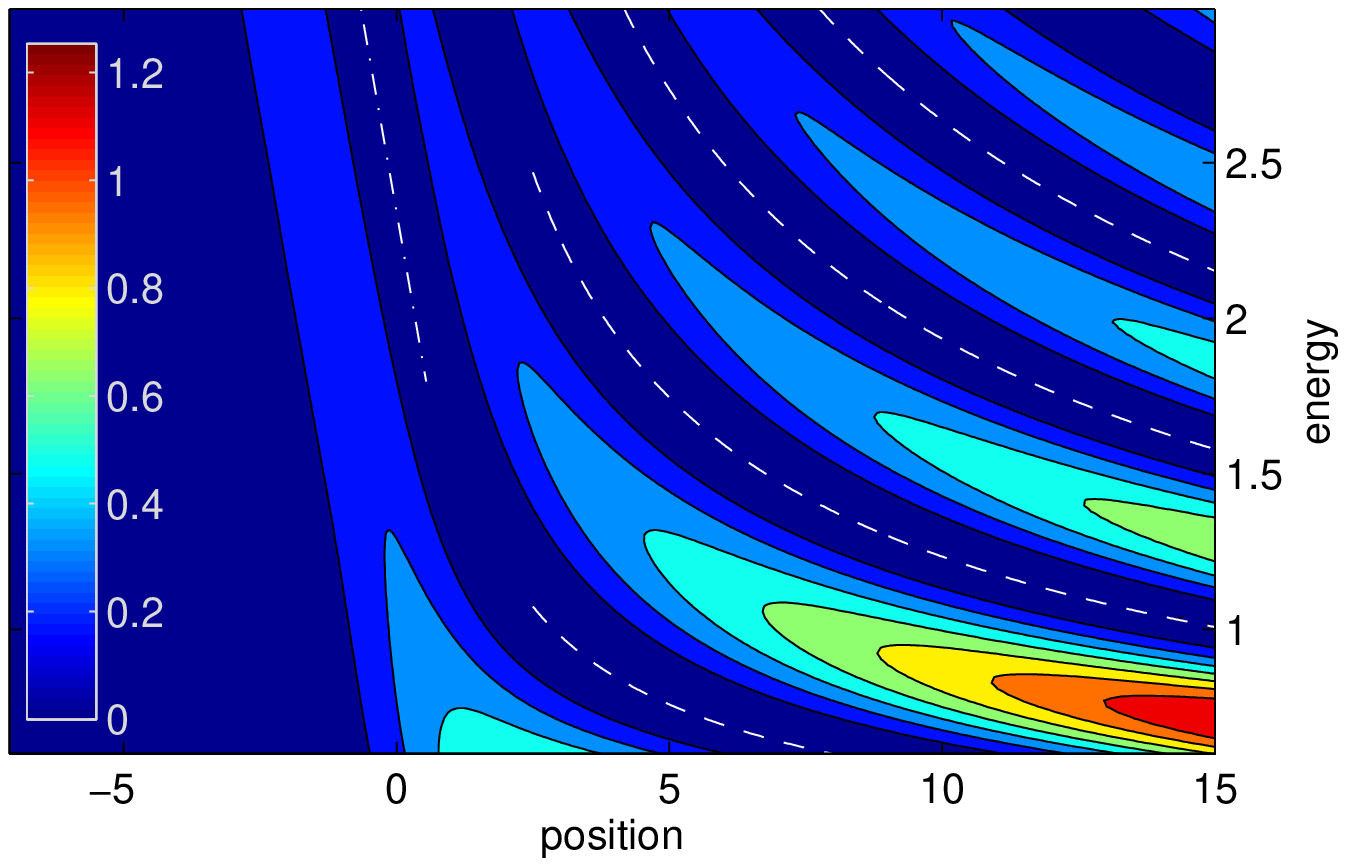} 
\\
   \includegraphics*[width=6.5cm]{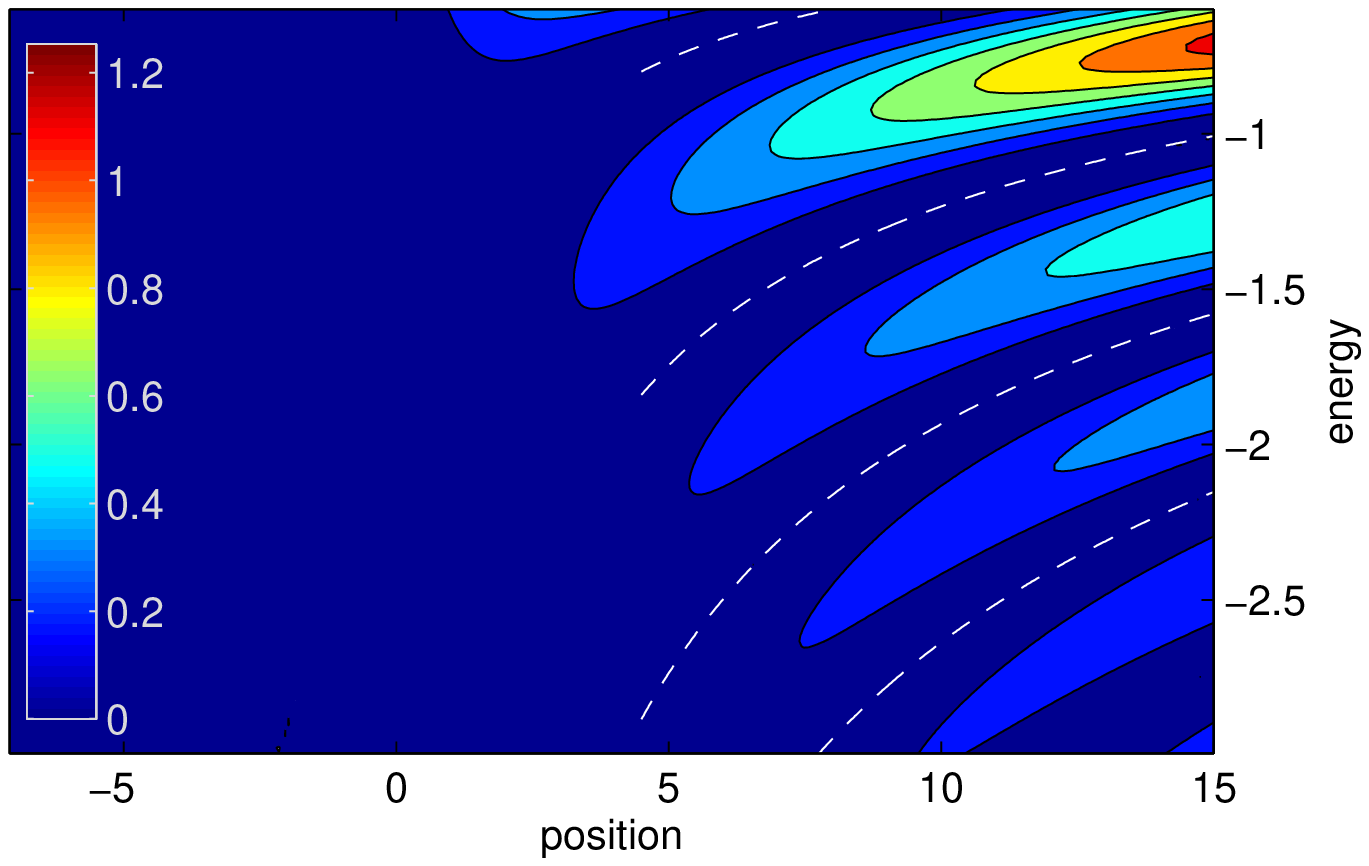} 
\end{tabular}
&
   \raisebox{-27mm}{\includegraphics*[width=7.5cm]{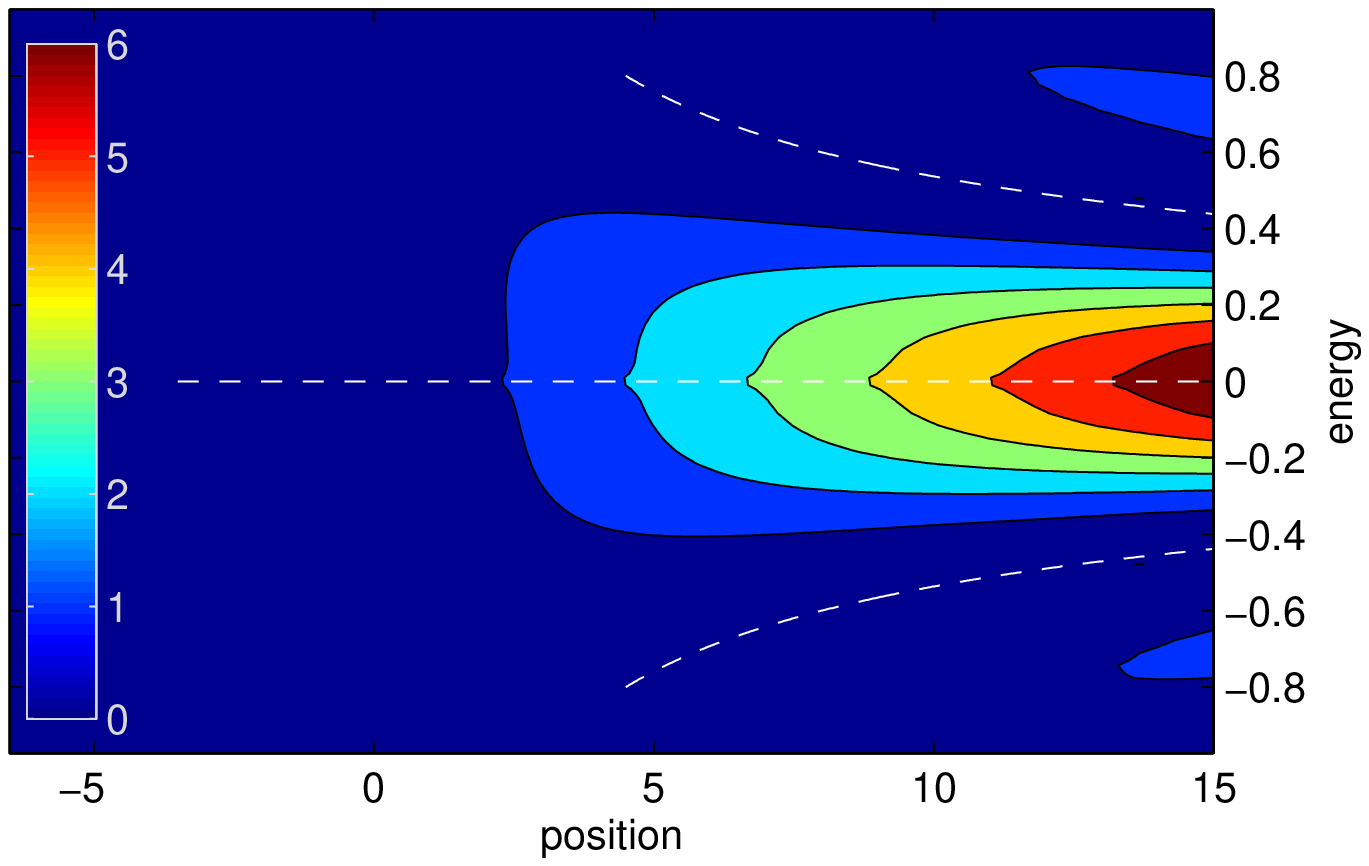}}
\end{tabular}
   \caption[]{Spectral representation of the field correlation 
   function $G(x, \tau )$. (\emph{left column}) Focus
   on positive energies: `particle mode'
   $u_E^2( z )$; and negative energies: `hole mode'
   $v_E^2( z )$. (\emph{right}) 
   Zoom into the low-energy region; note the change in color scale.
   Dashed and dash-dotted white: nodal lines explained in the text.} 
   \label{fig:field-spectrum}
\end{figure}
We show in Fig.\ref{fig:field-spectrum} a contour plot of two terms: 
the `particle spectrum' $u_E^2( z )$ and the
`hole spectrum' $v_E^2( z )$ (with the energy scale flipped). 
We recognise in the upper left quadrant (particles outside
the condensate) a straight nodal line $E \approx 2.34 - z$ (dash-dotted) 
that is characteristic for the 
Airy function $u_E( z ) \sim {\rm Ai}( - E - z )$.
As the modes enter the condensate, the nodal lines shift to
the pattern $E \sqrt{ 2 z } \approx 2.405,\, 5.520 \ldots$, the
first few roots of the Bessel function $J_0$ (dashed). This 
approximation is based on the asymptotic form~(\ref{eq:dense-with-phase})
and works well because the Bogoliubov phase shift is small:
$|\delta( E )| \ll \pi/2$ (see \citet{Diallo-phase}).

At low energies (bright central region 
in Fig.\ref{fig:field-spectrum}(\emph{right})), the spectra for
both particles and holes
converge to the same limit (see Eq.(\ref{eq:low-E-tilde-u}))
that is essentially given by the condensate density 
(see Sec.\ref{s:low-energies}).
The occupation number $\bar N \sim T / E$ 
in Eq.(\ref{eq:spectrum-field-corr}) thus leads to an infrared
divergence of the average density $n( z ) = G(z, 0)$.
This has been regularised by introducing the quasi-condensate concept
\citep{Kagan00,Andersen02c}: the divergence mainly arises from
phase fluctuations which can be subtracted. See the discussion of
the density correlations below.

Returning to the hole mode $v_E( z )$ 
{}(Fig.\ref{fig:field-spectrum}(\emph{bottom left})), we see
that it is confined to the dense region
and follows similar nodal lines as $u_E( z )$ as expected
from the boundary condition Eq.(\ref{eq:dense-with-phase}).
The contour plot provides a representation of the so-called depletion density
\begin{equation}
n_d( z ) = 
\int\limits_{0}^{\infty}\!\frac{ {\rm d}E }{ \pi } v_E^2( z )
\label{eq:def-depletion}
\end{equation}
which is simply the zero-temperature limit of the
non-condensate density
in Eq.(\ref{eq:spectrum-field-corr}) 
(a measure of quantum fluctuations). 
We have checked that this
integral matches in the dense region (i.e., $E \sqrt{ 2 z } \gg 1$) 
with the corresponding
result for a homogeneous system
\footnote{A useful parametrisation for the Bogoliubov amplitudes 
in a homogeneous system is
$u_k = \cosh(\eta_k/2)$,
$v_k = -\sinh(\eta_k/2)$ with $\sinh \eta_k = \phi^2 / E_k$. 
The dispersion relation is 
$E_k^2 = k^4 + 2 k^2 \phi^2$.
Therefore in the dense limit $\phi^2 \gg E_k$:
$v^2_k \approx {\rm e}^{ \eta_k }/ 4 \approx \phi^2 / 2 E_k$.
}
(local density approximation
with constant condensate $\phi$)
\begin{equation}
n_{d, {\rm LDA}} = 
\int\limits_{-\infty}^{\infty}\!\frac{ {\rm d}k }{ 2\pi }
v_k^2 \approx 
\phi^2
\int\limits_{0}^{\infty}\!\frac{ {\rm d}k }{ 2\pi E_k }
\label{eq:depletion-homogeneous}
\end{equation}
where the modes are labelled by the wave vector $k$ and the
dispersion relation is approximately linear $E_k \approx \sqrt{ 2 }\, k \phi$.
The logarithmic infrared divergence can also be cured with suitable
subtractions \citep{Andersen02c,AlKhawaja02b,Mora03}.

\subsection{Density fluctuation modes}
\label{s:density-fluctuations}

As a second application, we consider the leading order Bogoliubov
contribution to the dynamic density correlations
\begin{equation}
S( x, y, t - t' ) = 
{\textstyle\frac12} \langle \{
\rho( x, t ) 
,\,
\rho( y, t' ) 
\}
\rangle
- 
n(x) n(y)
\label{eq:}
\end{equation}
The curly brackets denote a symmetrised operator product 
for the particle density $\rho( z, t ) = \psi^\dag( z, t ) \psi( z, t ) $.
Its average $n( z ) = \langle \rho( z, t ) \rangle$ 
does not depend on time (see Eq.(\ref{eq:spectrum-field-corr})
for $\tau = 0$). The expectation value is worked
out using the Bogoliubov shift~(\ref{eq:expand-field})
and expressed in
terms of the occupation numbers~(\ref{eq:thermal-expectation}),
using the Wick theorem (gaussian statistics). Our result
is consistent with Eq.(52) of \citet{Eckart08} 
and reads
\begin{eqnarray}
&& S( x, y, \tau ) =
\mathop{\rm Re}\left\{ 
	G( x, y, -\tau ) \Delta( x, y, \tau )
\right\}
\nonumber\\
&& \quad
{} +
\phi(x ) \phi( y ) 
\int\limits_{0}^{\infty}\!\frac{ {\rm d}E }{ \pi }
\cos( E \tau )
\left\{
f_E( x ) f_E( y ) 
\bar N(E)
\right.
\nonumber\\
&& {} \left. \qquad
+
v_E( x ) f_E( y ) 
+
\left( x \leftrightarrow y \right)
\right\}
	\nonumber\\
&& \quad 
{} + 
\mbox{4th order terms}
\label{eq:result-dens-corr}
\end{eqnarray}
where the first line involves the correlation function 
of Eq.(\ref{eq:def-field-corr})
and the field commutator
\begin{eqnarray}
\Delta( x, y, t - t' ) &=& 
\left[ \psi( x, t ) ,\, \psi^\dag( y, t' ) \right]
	\nonumber\\
&=&
\int\limits_{0}^{\infty}\!\frac{ {\rm d}E }{ \pi }
\left\{
u_E( x ) u_E( y ) \, {\rm e}^{ - {\rm i} E (t - t') }
\right.
\nonumber\\
&& \qquad {} \left.
-
v_E( x ) v_E( y ) \, {\rm e}^{ {\rm i} E (t - t') }
\right\}
\label{eq:}
\end{eqnarray}
Due to the completeness relation of the \BdG{} modes,
this goes over into $\delta( x - y )$ when $t \to 0$
{}(see~\ref{a:Wronskian}).
In Eq.(\ref{eq:result-dens-corr}), we use the `sum mode function'
\begin{equation}
f_E( z ) = u_E( z ) + v_E( z )
\label{eq:}
\end{equation}
which is, by a property of the \BdG{} equations, orthogonal
to the condensate $\phi( z )$ with respect to the scalar 
product~(\ref{eq:def-L2-norm}).
The `4th order terms' of the last line arise from products 
of four Bogoliubov operators $a_E$ and $a_E^\dag$.
Note that the second line features, for $x = y$, an integral that
is similar to infrared-regularised thermal densities introduced by
\citet{Andersen02c,AlKhawaja02b,Mora03}.
This illustrates the consistency of these procedures,
since their goal is to eliminate from the density
spurious contributions attributed to phase fluctuations.

\begin{figure}[htbp]
\begin{center}
\includegraphics*[width=7.5cm]{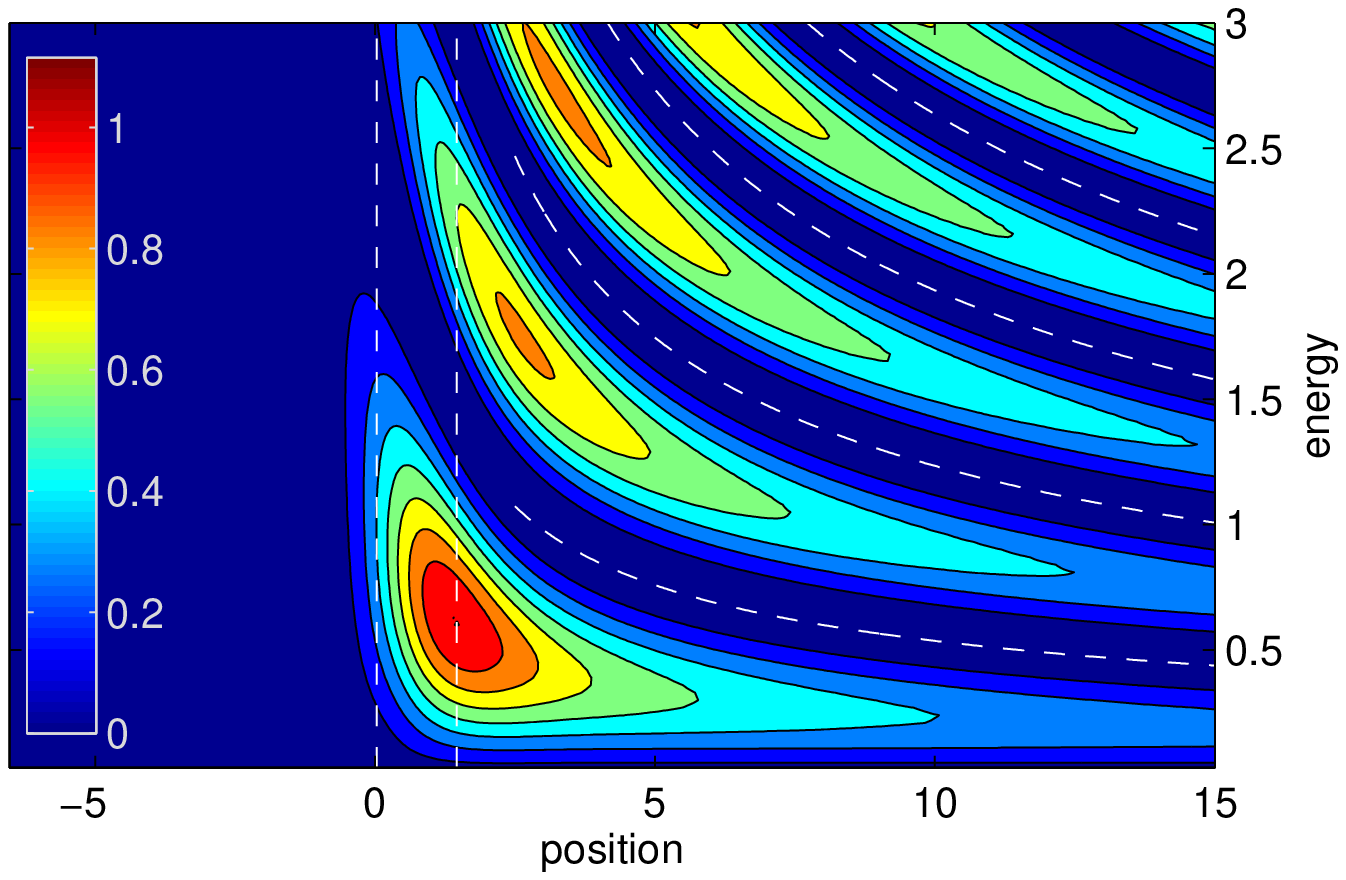}
\includegraphics*[width=7.5cm]{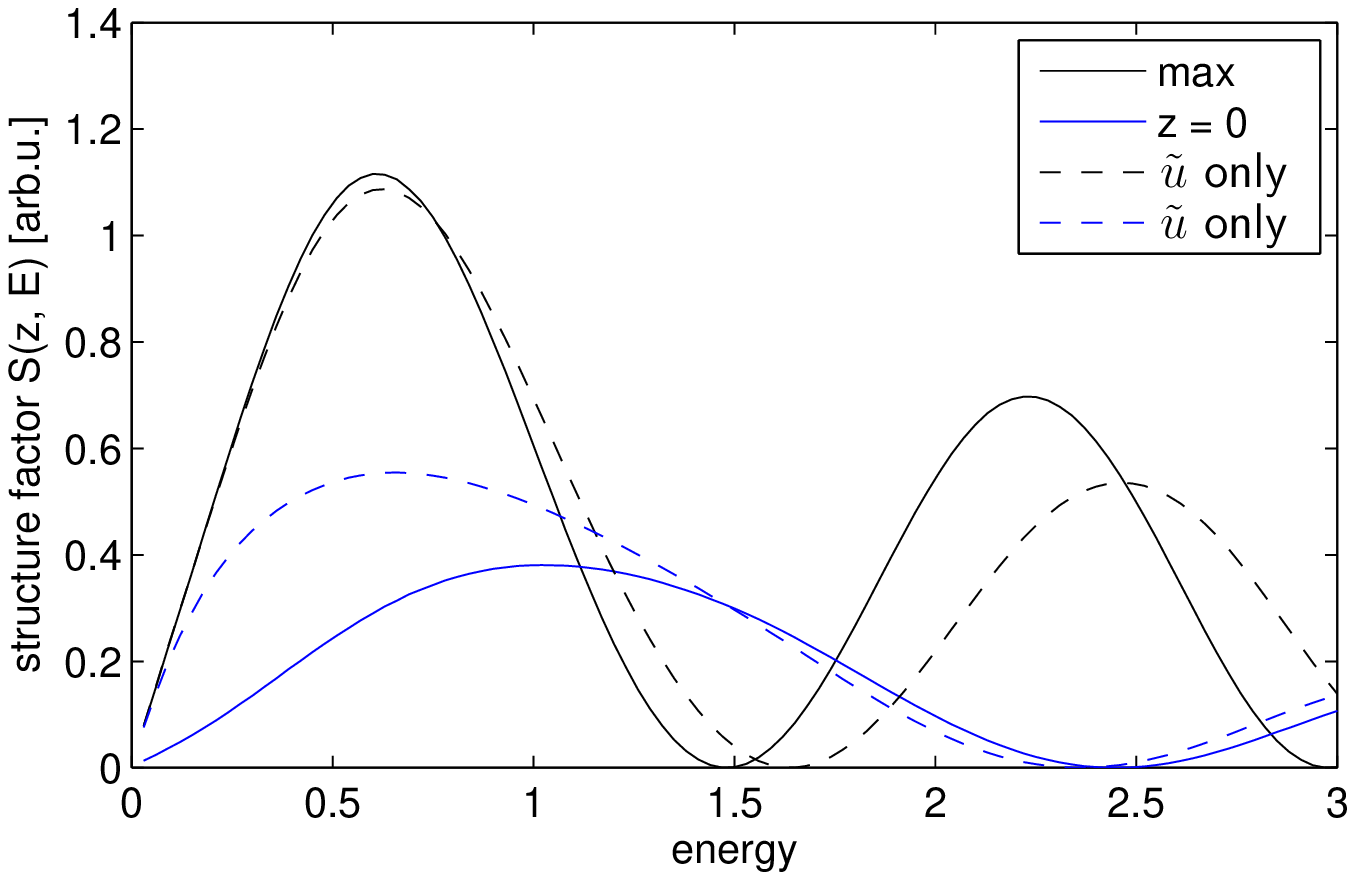}
\end{center}
\caption[]{Spectrum of local density fluctuations
due to Bogoliubov excitations in thermal equilibrium.
We plot Eq.(\ref{eq:beat-note-dens-spec})
which is the essentially the Fourier transform
of the second line in Eq.(\ref{eq:result-dens-corr})
for temperature $T = 1$ (using the units specified in 
Table~\ref{t:units}). This can be understood as a local
dynamic structure factor; it is symmetric in energy (only
$E \ge 0$ is shown).
(\emph{left})
Contour plot with nodal lines (dashed) as in Fig.\ref{fig:field-spectrum}.
(\emph{right})
Cut through the positions 
$z \approx 1$ (peak value of structure factor) and
$z = 0$ (border of Thomas-Fermi condensate).
Dashed: Bogoliubov
amplitudes calculated from `phase mode' $\tilde u$ only, `density
mode' $\tilde v$ omitted.
}
\label{fig:dens-spec}
\end{figure}

We focus for illustration purposes on the `beating' between the 
condensate and the elementary excitations and show 
in Fig.\ref{fig:dens-spec} the local spectrum
\begin{equation}
S_{\rm beat}( z, E ) = 
\frac{ \phi^2(z ) }{ \pi }
f_E^2( z )
( 2 \bar N(E) + 1 )
\label{eq:beat-note-dens-spec}
\end{equation}
We find this formula by including the part $\phi^2(z) \Delta(z, E)$ 
of the first line in Eq.(\ref{eq:result-dens-corr})
that is proportional to the condensate density.
The contour plot shows that the density
fluctuations are peaking near the condensate border. This is as
expected because deep inside a (quasi)condensate,
such fluctuations are penalised by the self-interaction energy.

The density fluctuation spectrum does not show any 
infrared divergence because for small $E$,
the sum mode $f_E( z )$ behaves proportional 
to (adiabatic angle $\theta \to \pi / 2$ in Eq.(\ref{eq:def-rotation}))
\begin{eqnarray}
&&
\left(
\cos{\textstyle\frac12}\theta - \sin{\textstyle\frac12}\theta
\right)
\tilde u_E
\approx
\frac{ E }{ \sqrt{ 2 } \, \phi^2 }
\tilde u_E
\nonumber\\
&&
\mbox{for } \phi^2( z ) \gg E
\label{eq:phase-fluct-suppressed}
\end{eqnarray}
The scaling linear in $E$ at finite temperature can be seen
in Fig.\ref{fig:dens-spec}(\emph{right}). 
This plot also illustrates that 
the `trapped mode' $\tilde v_E( z )$ which is 
localised near the border (Fig.\ref{fig:u-waves}(\emph{right})),
gives a significant contribution
{}(compare dashed and solid lines in Fig.\ref{fig:dens-spec}(\emph{right})).
The comparison yields the interesting result that the beating between 
this mode and the condensate is 
actually \emph{reducing} rather than enhancing low-frequency 
density fluctuations (set of lower curves for $z = 0$).

\section{Conclusion}

The elementary excitations of a Bose condensate 
(Bogoliubov spectrum) are well-known
in a homogeneous system~\citep{PitaevskiiStringari} and also
within some approximations for a harmonically trapped 
gas~\citep{AlKhawaja02b,Stringari96,Oehberg97d,Stringari98}.
We have analysed in this paper the border region where the
condensate density smoothly goes to zero, providing a detailed
look at the physics beyond the Thomas-Fermi approximation.
Previous work has focused on the condensate kinetic 
energy \citep{Dalfovo96,Lundh97}, ignoring the contribution
of elementary excitation, and on the stability with respect to
vortex formation, taking into account motion parallel to 
the border of the condensate \citep{Lundh97,Anglin01b}.
The mode functions provided here typically extend into the
bulk of the condensate and would correspond in a three-dimensional
isotropic trap to radially symmetric (angular momentum $l = 0$) modes.
Our main result is that the gradient in the
condensate density couples elementary excitations that
are mainly `phase-like' and `density-like', an effect clearly
beyond the local-density approximation. This leads to 
density fluctuation modes that are localised near the border of
the condensate. These fluctuations may be detected with scattering
experiments using a focused probe beam that probe the dynamic structure 
factor locally, similar to the setup of \citet{Onofrio00}.
Alternatively, one may directly analyse density-density 
correlations when an elongated system is imaged. This
may be complemented by launching,
with a suitable pulse sequence, 
an elementary excitation coming from the bulk (dense quasi-condensate), 
similar to the suggestion of \citet{Brunello00}.
We also believe that the methods developed here 
provide a stepping stone towards a self-consistent description
of an inhomogeneous Bose gas at finite temperature, using for example
the modified Popov theory of \citet{Andersen02c} or the Bogoliubov theory
for quasi-condensates of \citet{Mora03}.
The border region
where the density drops is particularly interesting here because of
the possibility of entering a strongly correlated phase, see for
example \citet{Trebbia06a,Jacqmin11}, and \citet{Vogler13}.

\paragraph{Acknowledgements.}
This work has been supported by a final-year student grant 
awarded to A.D by Universit\"at Potsdam.

\appendix

\section{Wronskians and normalisation}
\label{a:Wronskian}

We start by a generalisation of the Wronskian for the \BdG{} problem 
(two coupled equations). Our proposed definition is
\begin{eqnarray}
&&
W[u, v, u_1, v_1] = 
W[ u, u_1 ] + W[ v, v_1] 
\nonumber\\
&&
= 
u u_1' - u_1 u' + v v_1' - v_1 v'
\label{eq:BdG-Wronskian}
\end{eqnarray}
where the prime denotes the first derivative.
We assume that all modes including the condensate $\phi$ are real.
The advantage of this combination are the following manipulations 
that can be applied to the pair of \BdG{} equations
\begin{eqnarray}
E u &=& - u'' + H u + \phi^2 v
	\label{eq:def-BdG-simple-1}
\\
- E v &=& - v'' + H v + \phi^2 u 
	\label{eq:def-BdG-simple-2}
\end{eqnarray}
where $H = V + 2 |\phi|^2 - \mu$ is the Hartree-Fock potential.
Consider another pair of solutions $u_1$, $v_1$ that solves
the same equations with energy eigenvalue $E_1$.
Multiply Eq.(\ref{eq:def-BdG-simple-1}) with $u_1$, and 
Eq.(\ref{eq:def-BdG-simple-2}) with $v_1$,
take the sum, and subtract the corresponding equation for $u_1$ 
multiplied by $u$ etc. On the left-hand side, we get
$( E - E_1 ) (u_1 u - v_1 v)$, proportional to the integrand of 
the generalised
${\rm L}\!^2$-scalar product
in the \BdG{} space. On the right-hand side, we find the derivative
of the Wronskian $W[u, v, u_1, v_1]$: the Hartree potential drops 
out as in the Schr\"odinger equation; also the coupling terms involving
the condensate are cancelled:
$\phi^2 ( u_1 v + v_1 u - u v_1 - v u_1) = 0$. 
Integrating and using the boundary conditions~(\ref{eq:dilute-asymptote}) 
for $z \to -\infty$, 
we get
\begin{equation}
\int\!{\rm d}z\left[
u_1( z ) u( z ) - v_1( z ) v( z ) 
\right] = 
\lim_{z \to \infty}
\frac{ W[ u, v, u_1, v_1] }{ E - E_1 }
	\label{eq:link-L2-and-Wronskian}
\end{equation}
In other words: the scalar product can be analyzed \emph{locally}
from the asymptotic behavior of the mode functions. The orthogonality
between modes with different energies in the discrete spectrum follows
immediately (the Wronskian vanishes at both ends). 

We continue by analyzing the continuous spectrum for the linear 
potential.
Recall the asymptotic form deep in the condensate
from Eq.(\ref{eq:dense-with-phase}):
\begin{eqnarray}
&&
z \to +\infty:
\qquad
\nonumber\\
&&
u( z ) \to 
A \frac{ (2 z)^{1/4} }{ \sqrt{ E } } \cos( E \sqrt{2 z} - \pi/4 + \delta)
\label{eq:recap-normalized-u}
\end{eqnarray}
and similarly for $v(z)$ with amplitude $B$. In this limit,
the two amplitudes are given by the rotation back from the
adiabatic basis
\begin{eqnarray}
&&
\left( \begin{array}{c} A \\ B \end{array} \right)
=
\left( \begin{array}{c} {\cal A} \cos \theta/2 
	\\ - {\cal A} \sin\theta/2 \end{array} \right)
\,,\qquad
\nonumber\\
&&
\tan\theta = \frac{ \phi^2( z ) }{ E }
\label{eq:def-rotated-basis}
\end{eqnarray}
where we have used that only the adiabatic mode $\tilde u$ 
`survives' and has amplitude ${\cal A}$. 
{}($\tilde v$ is localised in the Hartree-Fock-like well
near the border, see Fig.\ref{fig:u-waves}(\emph{right}).)
The scattering phase $\delta$ is therefore the same 
for both modes $u$ and $v$.
Similar amplitudes $A_{1}$ and
$B_{1}$ apply for the other solution at energy $E_1$.
The Wronskian then becomes 
(denoting $\varphi = E \sqrt{ 2 z } - \pi / 4 + \delta$ 
and similarly for $\varphi_1$)
\begin{eqnarray}
&&
\lim_{z \to \infty}
\frac{ W[ u, v, u_1, v_1] }{ E - E_1 }
=
\nonumber\\
&&
\frac{ A A_1 + B B_1 }{ E - E_1 } 
\Big(
\sqrt{ \frac{ E_1 }{ E } }
\cos \varphi \sin \varphi_1
-
\sqrt{ \frac{ E }{ E_1 } }
\sin \varphi \cos \varphi_1 
\Big)
\label{eq:}
\end{eqnarray}
{}Contributions from the derivatives $dA / dz , dB / dz$ 
would vanish like $1/z^{3/2}$ relative to this term, 
see Eq.(\ref{eq:TF-for-thetap}).
It is natural to interpret this as distribution with respect 
to the energies $E, E_1$, to be put under an integral.
The trigonometric functions can be re-written into
$\sin( \varphi + \varphi_1 )$,
this gives the expression
\begin{equation}
\frac{ A A_1 + B B_1 }{ 2 \sqrt{ E E_1 } }
\sin[ (E + E_1) \sqrt{ 2 z } - \pi/2 + \delta + \delta_1 ]
\label{eq:}
\end{equation}
Since both energies are positive, this is an oscillating function
as $z \to \infty$. It averages to zero 
if integrated over some interval $\Delta E > 2\pi/\sqrt{ 2 z }$
and therefore vanishes in the distribution sense.
The Wronskian is thus given by the phase difference term
$\sim \sin( \varphi - \varphi_1 )$ 
\begin{eqnarray}
&&
\lim_{z \to \infty}
\frac{ W[ u, v, u_1, v_1] }{ E - E_1 } 
\nonumber\\
&& =
( A A_1 + B B_1) \frac{ E + E_1 }{ 2 \sqrt{ E E_1 } }
\frac{ 
	\sin[ (E - E_1) \sqrt{ 2 z } + \delta - \delta_1 ]
	}{
	E - E_1 
	}
\nonumber
\\
	&&=
( A^2 + B^2 ) \pi
\delta( E - E_1)
\end{eqnarray}
where we recognised in the last fraction an oscillatory 
representation of a $\delta$-function
\begin{equation}
\lim_{t \to \infty} \frac{ \sin ( x t ) }{ x } = \pi \delta( x )
\label{eq:delta-representation}
\end{equation}
and used the continuity in energy of the phase shift
and of the amplitudes.
The latter sum to $A^2 + B^2 = {\cal A}^2$ and 
in Eq.(\ref{eq:dense-with-phase}), we chose the 
normalisation ${\cal A} = 1$. This
leads from Eq.(\ref{eq:link-L2-and-Wronskian}) 
to the orthogonality
\begin{equation}
\int\!\frac{ {\rm d}z }{ \pi } \left(
u_1( z ) u( z ) - v_1( z ) v( z ) 
\right) = 
\delta( E - E_1 )
	\label{eq:completeness-uv}
\end{equation}
which is the main result of this appendix. The symmetry transformation
$u \leftrightarrow v$ and $E \leftrightarrow -E$ of the \BdG{} 
problem~(\ref{eq:def-BdG-simple-1}, \ref{eq:def-BdG-simple-2}) 
gives the additional orthogonality relation
\begin{equation}
\int\!\frac{ {\rm d}z }{ \pi } \left(
u_1( z ) v( z ) - v_1( z ) u( z ) 
\right) = 0
	\label{eq:orthogonal-uv}
\end{equation}
As a consequence, we can easily check that the equations
\begin{eqnarray}
a(E) &=&
\int\!\frac{ {\rm d}z }{ \sqrt{ \pi } }
\left(
\psi( z ) u_E( z ) - \psi^\dag( z ) v_E( z )
\right)
\nonumber
\\
\psi( z ) &=& \int\limits_{0}^{\infty}\!\frac{ {\rm d}E }{ \sqrt{ \pi } }
\left(
a(E) u_E( z ) + a^\dag( E) v_E( z )
\right)
\label{eq:extract-mode-operator}
\end{eqnarray}
translate the standard commutation relation of the field operator
$[ \psi( x ), \psi^\dag( y ) ] = \delta( x - y )$
into an implementation of the canonical commutation relations for 
the elementary mode operators
\begin{equation}
[a( E), a^\dag( E') ] = \delta( E - E' )
\label{eq:}
\end{equation}
provided the mode functions $u$ and $v$ are normalised 
as in Eq.(\ref{eq:recap-normalized-u}) with ${\cal A} = 1$.
The Bogoliubov shift (condensate in Eq.(\ref{eq:expand-field}))
does not change this conclusion.
For a discussion of the zero mode in the \BdG{} problem and
the corresponding operators, see for example~\citet{Mora03}.

\section{Solving the \BdG{} equations numerically}
\label{a:numerics}

We use a standard differential equation solver for the 
Painlev\'e transcendent (Gross-Pitaevskii 
equation~(\ref{eq:GPE})). The solution that connects to the
Thomas-Fermi profile is actually numerically unstable, and
we match it around $z \sim 3$ with the asymptotic 
expansion~(\ref{eq:dense-asymptote}), keeping typically 
three terms.

To solve the \BdG{} equation~(\ref{eq:BdG-adiabatic-u}) 
in the open channel (mode $\tilde u$) in the adiabatic approximation, 
a standard
forward solver is used: initialise with the tunnelling 
asymptote~(\ref{eq:dilute-asymptote}) and check that the 
potential is linear there. Integrate forward until a position 
$z_1 \gg 1$ and match to a solution of the modified Coulomb
problem~(\ref{eq:modified-Coulomb})
\begin{equation}
\tilde u( z ) = \alpha( z ) j( z ) - \beta( z ) y( z )
	\label{eq:match-tilde-u}
\end{equation}
The coefficients $\alpha, \beta$ are conveniently calculated
with the help of the Wronskians
\begin{equation}
\alpha( z_1 ) = W[\tilde u, y]({z_1})
\,,
\qquad
\beta( z_1 ) = W[\tilde u, j]({z_1})
	\label{eq:Wronskians-for-matching}
\end{equation}
using the normalised Bessel-Coulomb solutions 
defined in Eq.(\ref{eq:Bessel-Coulomb-solutions}).

At the position $z_1$, the open potential $- \tilde k^2( z )$ 
may not yet have reached its
Coulomb asymptote $V_C( z )$ (see Eq.(\ref{eq:Coulomb-tail})), 
therefore the coefficients $\alpha, \beta$
are still slowly varying. The Wronskians~(\ref{eq:Wronskians-for-matching})
satisfy a first-order differential equation that can be derived using the 
procedure explained after Eq.(\ref{eq:def-BdG-simple-2}). This yields, 
for example,
\begin{eqnarray}
&&
\beta = \lim_{z \to \infty} W[\tilde u, j] 
\nonumber\\
&& = 
\beta( z_1 ) +
\int\limits_{z_1}^{\infty}\!
{\rm d}z (- \tilde k^2( z ) - V_C( z )) \tilde u( z ) j( z )
	\label{eq:extrapolate-Wronskian}
\end{eqnarray}
We choose the position $z_1$ such that the following approximation 
to the open potential is accurate enough
\begin{eqnarray}
&&
z \ge z_1: \qquad
\nonumber\\
&& 
- \tilde k^2( z ) - V_C( z ) 
\approx
- \frac{ E^4 / (4 z) }{ z^2 + z (z^2 + E^2)^{1/2} + E^2/2 }
	\label{eq:next-order-potential}
\end{eqnarray}
This term arises from the expansion of the root $(\phi^4 + E^2)^{1/2}$;
other contributions
(post-Thomas-Fermi correction, geometric potential)
are smaller. We find that for $z_1 \approx 15$, the relative error
is smaller than $10^{-3}$ for a wide range of energies.
We compute the integral~(\ref{eq:extrapolate-Wronskian})
numerically with the approximation~(\ref{eq:match-tilde-u}) for
$\tilde u$. It converges because the potential difference scales 
like $1/z^3$.
One can avoid the evaluation of oscillatory Bessel functions 
for large arguments by (i) using their asymptotic form and
(ii) shifting the integration contour into the complex plane
after some point $z \gtrsim \max\{ z_1, 2 E \}$ on the real axis. 
In this way,
one is keeping clear of the branch cut
of Eq.(\ref{eq:next-order-potential})
at $z = \pm {\rm i} E$.
This procedure now yields the extrapolated coefficients 
$\alpha, \beta$. The normalisation factor
for the wave function $\tilde u$ is then $1/(\alpha^2 + \beta^2)^{1/2}$,
and the scattering phase shift follows from
$\tan \delta = \beta / \alpha$.

The calculation of the density mode $\tilde v$ is based on
the adiabatic approximation for the inhomogeneous Schr\"odinger
equation~(\ref{eq:BdG-adiabatic}). We represent the
differential operator in the closed potential $\tilde \kappa^2( z )$
on a grid with a finite difference scheme. The size of the grid is
adapted to the support of the source term $\Lop \tilde u$.
Due to the nonzero minimum of the potential, 
zero energy is not in the spectrum of the differential operator,
hence the inhomogeneous equation is solved by a straightforward
matrix inversion.

\section{Trapped states}
\label{a:trap-spectrum}

The closed potential $\tilde \kappa^2( z )$ has linear asymptotes
on both sides (see its Thomas-Fermi 
approximation in 
Eqs.(\ref{eq:TF-closed-dilute}, \ref{eq:TF-closed-dense}) below).
Physically allowed eigenmodes therefore join into tunnelling solutions and
occur only for discrete eigenvalues $\epsilon_n$ (see 
Eq.(\ref{eq:closed-eigenstates}), not to be
confused with $E$ which remains a continuous parameter).

\begin{figure}[htbp]
\begin{center}
	\includegraphics*[width=8.5cm]{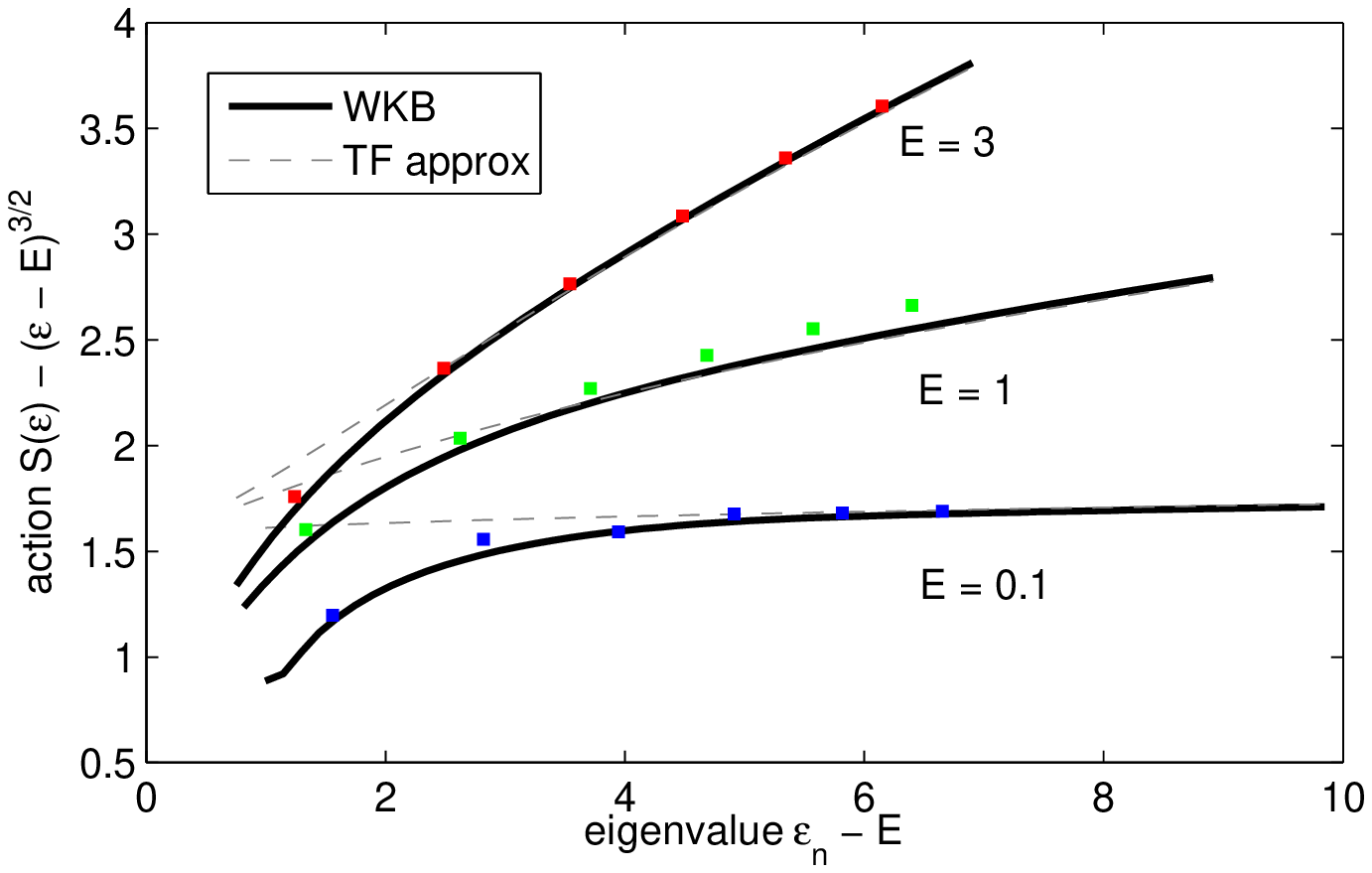}
\end{center}
   \caption[]{Spectrum $\{ \epsilon_n \}$ of trapped states (colored dots)
   compared to the Bohr-Sommerfeld rule~(\ref{eq:Sommerfeld-rule})
    (thick black line).
   The dashed line gives the analytical 
   approximation~(\ref{eq:result-Bohr-Sommerfeld-trapped-action}).
   To enhance the difference,
   the leading term $(\epsilon - E)^{3/2}$ has been subtracted
   from the action ($y$-axis).}
   \label{fig:trapped_wkb_BS}
\end{figure}

For the numerical calculation of the trapped states in the closed potential,
we use the finite-difference scheme of the preceding~\ref{a:numerics}
and take a standard sparse eigenvector solver. Examples are 
shown in Fig.\ref{fig:trapped-modes}(\emph{left column}). 
A comparison of the spectrum $\{ \epsilon_n \}$ with the familiar
Bohr-Sommerfeld quantisation rule is given in Fig.\ref{fig:trapped_wkb_BS}.
Recall that this rule is based on the action integral
\begin{eqnarray}
S( \epsilon ) &=& \frac{ \pi }{ 2 } 
+ \int\limits_{z_1}^{z_2}\!{\rm d}z \, p( z; \epsilon )
\,,\qquad
\nonumber\\
p( z; \epsilon ) &=& \sqrt{ \epsilon - \tilde \kappa^2( z ) }
	\label{eq:action}
\end{eqnarray}
where $z_{1,2}$ are the left and right roots of
$p^2( z; \epsilon )$ (also known as turning points).
The phase $\pi / 2$ arises from the Langer correction 
at both turning points~\citep{MessiahI}. 
The eigenvalues are then approximately given by
\begin{equation}
S( \epsilon ) = \pi ( n + 1 )
\,,\qquad
n = 0, 1, 2, \ldots 
	\label{eq:Sommerfeld-rule}
\end{equation}
The action integral,
computed numerically, is plotted as thick lines in 
Fig.\ref{fig:trapped_wkb_BS}, and a good agreement with the numerically
computed eigenvalues is found. 
For the plot, the colored squares mark the 
pair $(\epsilon_n, \pi ( n + 1 ))$. To enhance the difference, we
have subtracted the leading term $(\epsilon - E)^{3/2}$ from the
action, see Eq.(\ref{eq:result-Bohr-Sommerfeld-trapped-action}) below.

The dashed lines in the figure show the Thomas-Fermi approximation
to the action that can be computed analytically and provides a
relatively accurate estimate. The closed potential is approximated
by 
\begin{eqnarray}
z \le 0: &&
\tilde \kappa^2 \approx E - z
	\label{eq:TF-closed-dilute}
\\
z \ge 0: &&
\tilde \kappa^2 \approx z + \sqrt{ E^2 + z^2 }
	\label{eq:TF-closed-dense}
\end{eqnarray}
These formulas are also useful to estimate the position of the left
and right turning points $z_{1,2}$. The action integral gives
$	\frac23 ( \epsilon - E )^{3/2}
$
from the region $z_1 \ldots 0$, and the range $0 \ldots z_2$ can
be evaluated with
the substitution $z = E \sinh t$. Summing the two, we get
\begin{eqnarray}
	S &=& (\epsilon - E)^{3/2} 
	+ \frac{\pi}{2}
	\nonumber\\
	&&
	+ \frac{ E }{ 2 } (\epsilon - E)^{1/2} 
	- \frac{ E^2 }{ 2 \sqrt{ \epsilon } }
	\mathop{\rm arctanh}\sqrt{\frac{ \epsilon - E }{ \epsilon } }
	\,,
	\label{eq:result-Bohr-Sommerfeld-trapped-action}
\end{eqnarray}
The first two terms give with the Bohr-Sommerfeld 
rule~(\ref{eq:Sommerfeld-rule}) 
the eigenvalue spectrum 
$\epsilon_n \sim E + [\pi (n + \frac12)]^{2/3}$
mentioned after 
Eq.(\ref{eq:solution-trapped-amplitudes}). The scaling law
$\epsilon_n \sim n^{2/3}$ illustrates the
non-equidistant spectrum in this anharmonic well.
Eq.(\ref{eq:result-Bohr-Sommerfeld-trapped-action}) captures
relatively well the numerically computed action (compare dashed
and solid lines in Fig.\ref{fig:trapped_wkb_BS}), except at low
energies where the Thomas-Fermi approximation fails to reproduce
the shape of the potential.

\end{document}